\documentclass[conference,compsoc]{IEEEtran}

\usepackage[hyphens]{url}
\usepackage[breaklinks]{hyperref}

\usepackage{enumitem}

\usepackage{graphicx}

\usepackage[scaled=0.88]{helvet}

\usepackage{mathtools}
\usepackage[linesnumbered,ruled,noend]{algorithm2e}
\usepackage[operators,sets,probability,landau,lambda,primitives]{cryptocode}
\renewcommand{\pccomment}[2][1em]{\hspace{#1}{\mbox{//\;}\text{\scriptsize#2}}}
\usepackage{amssymb}
\usepackage[a]{esvect} %

\usepackage[capitalise]{cleveref}
\usepackage{booktabs}

\usepackage{array}
\usepackage{multirow}
\usepackage{makecell}
\usepackage{xcolor,colortbl}
\usepackage{siunitx}
\usepackage{circledsteps}
\usepackage{tablefootnote}
\usepackage{threeparttable}

\usepackage{acro}
\acsetup{case-sensitive=false}
\DeclareAcronym{pprl}{short=PPRL, long=Privacy-Preserving Record Linkage}
\DeclareAcronym{lsh}{short=LSH, long=Locality-Sensitive Hashing}
\DeclareAcronym{smc}{short=SMC, long=Secure Multi-Party Computation}
\DeclareAcronym{he}{short=HE, long=Homomorphic Encryption}
\DeclareAcronym{psi}{short=PSI, long=Private Set Intersection}
\DeclareAcronym{fpsi}{short=FPSI, long=Fuzzy Private Set Intersection}
\DeclareAcronym{OT}{short=OT, long=Oblivious Transfer}
\DeclareAcronym{OTe}{short=OTe, long=Oblivious Transfer Extension}

\newcommand{\etal}{et~al.\xspace}

\def\addvalue#1#2{\expandafter\gdef\csname my@data@#1\endcsname{#2}}%
\def\usevalue#1{\csname my@data@#1\endcsname}

\newcommand{\reqdef}[3]{%
\vspace{0.2em}
\addvalue{#2}{RQ.#1}%
\hypertarget{#2}\noindent\textit{\usevalue{#2}: #3.}
}
\newcommand{\reqlinky}[1]{\hyperlink{#1}{\usevalue{#1}}}

\newcommand{\threshold}{\ensuremath{\tau}\xspace} 
\newcommand{\attributesim}[1]{\ensuremath{\textsf{sim}_{#1}}\xspace}
\newcommand{\attributesimeq}{\attributesim{=}}
\newcommand{\attributesimds}{\attributesim{\approx}\xspace}
\newcommand{\simjaccard}{\attributesim{J}\xspace}
\newcommand{\attributecount}{\ensuremath{n_a}\xspace}

\newcommand{\gramlen}{\ensuremath{q}\xspace}
\newcommand{\grams}[2]{\ensuremath{\textsf{grams}_{#1}(#2)}\xspace}
\newcommand{\qgrams}{\gramlen-grams\xspace}
\newcommand{\Mainzelliste}{\textsf{MainSEL}\xspace}
\newcommand{\EpiLink}{\textsf{EpiLink}\xspace}

\newcommand{\lshr}{\ensuremath{r}\xspace}
\newcommand{\lshb}{\ensuremath{b}\xspace}
\newcommand{\dimension}{\ensuremath{l}\xspace}
\newcommand{\distance}{\ensuremath{d}\xspace}
\newcommand{\hammingdist}{\ensuremath{\distance_H}\xspace}
\newcommand{\hammingweight}{\ensuremath{w_H}\xspace}
\newcommand{\range}[1]{\ensuremath{[#1]}\xspace}
\newcommand{\otfpsi}{\textsf{otFPSI}\xspace}
\newcommand{\otfpsihyb}{\textsf{otFPSI-h}\xspace}
\newcommand{\otfpsiss}{\textsf{otFPSI-ss}\xspace}
\newcommand{\otfpsisshyb}{\textsf{otFPSI-ss-h}\xspace}
\newcommand{\otfpsissb}{\textsf{otFPSI-ssb}\xspace}
\newcommand{\otfpsissbhyb}{\textsf{otFPSI-ssb-h}\xspace}

\newcommand{\recorduniverse}{\ensuremath{\mathcal{R}}\xspace}
\newcommand{\embedding}{\ensuremath{E}\xspace}

\newcommand{\descr}[1]{\noindent \textbf{#1}}
\newcommand{\subdescr}[1]{\noindent \emph{#1}}

\usepackage{pifont}

\newcommand{\modulus}{\ensuremath{p}\xspace}
\newcommand{\group}{\ensuremath{\ZZ_\modulus}\xspace}
\newcommand{\setA}{\ensuremath{Q}\xspace}
\newcommand{\sizeA}{{\ensuremath{n_Q}}\xspace}
\newcommand{\elementA}{\ensuremath{q}\xspace}
\newcommand{\setB}{\ensuremath{R}\xspace}
\newcommand{\sizeB}{{\ensuremath{n_R}}\xspace}
\newcommand{\elementB}{\ensuremath{r}\xspace}
\newcommand{\norgs}{{\ensuremath{n_T}}\xspace}
\newcommand{\partyA}{\ensuremath{\mathcal{Q}}\xspace}
\newcommand{\partyB}{\ensuremath{\mathcal{R}}\xspace}
\newcommand{\server}[1]{\ensuremath{\mathcal{S}_{#1}}\xspace}
\newcommand{\organization}[1]{\ensuremath{\mathcal{T}_{#1}}\xspace}
\newcommand{\registrationdb}[1]{\ensuremath{\mathcal{R}_{#1}}\xspace}
\newcommand{\shareA}[1]{\ensuremath{\overline{#1}\xspace}}
\newcommand{\shareB}[1]{\ensuremath{\widehat{#1}\xspace}}
\newcommand{\serverA}{\server{1}}
\newcommand{\serverB}{\server{2}}
\newcommand{\prfff}[2]{F_{#2}(z,#1)}
\newcommand{\ffpsi}{\ensuremath{\mathcal{F}_\text{FPSI}}\xspace}
\newcommand{\ffpsipart}[1]{\ensuremath{\mathcal{F}_{\text{FPSI},#1}}\xspace}
\newcommand{\fssfpsi}{\ensuremath{\mathcal{F}_\text{ssFPSI}}\xspace}
\newcommand{\fssfpsipart}[1]{\ensuremath{\mathcal{F}_{\text{ssFPSI},#1}}\xspace}
\newcommand{\protooutput}[1]{\ensuremath{\textsf{out}^{#1}}\xspace}
\newcommand{\protooutputpart}[2]{\ensuremath{\textsf{out}^{#1}_{#2}}\xspace}
\newcommand{\view}[2]{\ensuremath{\textsf{view}^{#1}_{#2}}\xspace}
\newcommand{\disteq}{\equiv}
\newcommand{\distindist}{\overset{c}{\disteq}}

\newcommand{\xor}{\oplus}

\newcommand{\name}{\textsf{xDup}\xspace}

\DeclareSIUnit \x {\texttimes}

\newcommand{\kwfor}{\textbf{for}\xspace}
\newcommand{\kwif}{\textbf{if}\xspace}
\newcommand{\kwreturn}{\textbf{return}\xspace}
\newcommand{\getsp}{\gets_\modulus}
\newcommand{\ind}{$\quad$}
\newcommand{\indd}{$\quad\quad$}
\newcommand{\inddd}{$\quad\quad\quad$}

\newcounter{linecounter}

\newcommand{\setlisting}[1]{\def\currentlisting{#1}}

\newcommand{\li}[1][]{%
    \ifx\relax#1\relax%
        {\tiny\thelinecounter}\;\stepcounter{linecounter}%
    \else%
        \expandafter\xdef\csname linenum@\currentlisting @#1\endcsname{\thelinecounter}%
        {\tiny\thelinecounter}\;\stepcounter{linecounter}%
    \fi%
}

\newcommand{\lineref}[2]{%
    \ifcsname linenum@#1@#2\endcsname%
        \csname linenum@#1@#2\endcsname%
    \else%
        \textbf{??}%
    \fi%
}

\ifCLASSOPTIONcompsoc
  \usepackage[nocompress]{cite}
\else
  \usepackage{cite}
\fi

\ifCLASSOPTIONcompsoc
  \usepackage[caption=false,font=footnotesize,labelfont=sf,textfont=sf]{subfig}
\else
  \usepackage[caption=false,font=footnotesize]{subfig}
\fi

\usepackage[full]{optional}

\begin{document}

\date{}

  \title{\name: Privacy-Preserving Deduplication for Humanitarian Organizations using Fuzzy PSI\\[2mm]
\footnotesize{\normalfont\emph{This is the full version of the conference paper published at the IEEE Symposium on
  Security and Privacy 2026.\\ This version includes extended appendices. Please cite the conference version.}}\vspace{-2mm}}

\author{
    \IEEEauthorblockN{Tim Rausch, Sylvain Chatel, Wouter Lueks}
    \IEEEauthorblockA{CISPA Helmholtz Center for Information Security, Saarbrücken, Germany
    \\\{tim.rausch, sylvain.chatel, lueks\}@cispa.de}
}

\maketitle

\begin{abstract}
Humanitarian organizations help to ensure people's livelihoods in crisis situations. 
Typically, multiple organizations operate in the same region. To ensure that the limited budget of these organizations can help as many people as possible, organizations perform cross-organizational deduplication to detect duplicate registrations and ensure recipients receive aid from at most one organization.
Current deduplication approaches risk privacy harm to vulnerable aid recipients by sharing their data with other organizations. We analyzed the needs of humanitarian organizations to identify the requirements for privacy-friendly cross-organizational deduplication fit for real-life humanitarian missions.
We present \name, a new practical deduplication system that meets the requirements of humanitarian organizations and is two orders of magnitude faster than current solutions. \name builds on Fuzzy PSI, and we present \otfpsi, a concretely efficient Fuzzy PSI protocol for Hamming Space without input assumptions. We show that it is more efficient than existing Fuzzy PSI protocols.
\end{abstract}

\section{Introduction}
\label{sec:introduction}
Humanitarian organizations assist people who have been affected by crises and situations caused by, for example, natural disasters, armed conflict, health crises, or famine. They support people's livelihoods by providing essential goods like food or hygiene items, and (health) services. Yet, the financial resources of these organizations are limited. Thus, they take measures to ensure that their limited resources can help as many people as possible.
Our conversations with humanitarian organizations highlighted \emph{deduplication}~\cite{Douglas2023b, haffar2022, DIGID, CALP} as a key measure. In crisis situations, typically many organizations are involved in assisting affected populations~\cite{ifrc2023,DIGIDblog}. As a result, aid recipients could -- accidentally or on purpose -- register with several organizations at once, potentially resulting in others not receiving the assistance they need~\cite{TTOI2011}. 
Duplicate registrations are estimated as high as \qty{15}{\percent}~\cite{ifrc2023, Douglas2023b}. A deduplication process enables organizations to check whether newly registered recipients are already registered with another organization and enables organizations to take action in these cases.
Any deduplication system must provide strong privacy guarantees as humanitarian aid recipients are an extremely vulnerable population~\cite{Wille2023b}. For recipients, refusing to receive aid is typically not an option -- yet they can suffer dire consequences when their privacy is not sufficiently safeguarded~\cite{hrw2021rohingya,ciesielski2022afghans}. 

Organizations can deduplicate recipients based on three categories of data: unique identifiers, biometrics, and biographical data \cite{ifrc2023}. Each approach has its unique challenges: 
Reliable unique identifiers like (government-issued) identity documents are often unavailable in the regions and settings humanitarian organizations operate in. 
Biometric data (e.g., fingerprints, iris scans) \cite{genkey2016abis,wfpscopeDedup} is inherently sensitive, and its collection can be seen as a substantial intrusion into recipients' privacy. 
Biographical data of recipients (e.g., name, date of birth, gender) is typically manually collected during registration and may be error-prone, requiring privately comparing inconsistent data~\cite{ifrc2023, haffar2023, DIGIDslides}.

Organizations increasingly use biographical data~\cite{Douglas2023b, DIGID, haffar2022b, haffar2023b} to abstain from the highly privacy-invasive collection of biometric data~\cite{icrcbiometrics,engineroom2023biometrics,oxfamengineroom2018biometrics,hpg2021biometrics,currion2015biometrics} and to avoid relying on externally issued unique identifiers. 

In this paper, we address the challenge of privacy-preserving deduplication based on biographical data by proposing \name, a new cross-organizational deduplication system. We elicit requirements for such a system from several discussions with organizations and a review of humanitarian publications, and tailor \name to these requirements:
(1) \emph{Privacy} of recipients must be ensured: No information about non-duplicates should be leaked to other organizations, protecting not only recipients but also NGOs. (2) The deduplication process must be fine-tuned to prioritize a low false-positive rate, so that recipients are not falsely flagged. False positives require manual handling, causing manual effort and leakage about non-duplicate registrations. 
(3) The system must \emph{scale}: Each organization records in the order of \num[exponent-mode=input]{100000} registrations~\cite{Edalatnejad2024, WFPBreport, WFPBB}, and may register thousands of new recipients per week \cite{UNHCRSudan,UNHCRUganda}.

Existing building blocks cannot satisfy these requirements. %
Fuzzy matching techniques~\cite{Lai2006, Schnell2009, Durham2010,Karakasidis2011,Schnell2011,Vatsalan2012b,Durham2014,Ranbaduge2014,Vatsalan2014} based on Bloom filters are efficient but susceptible to privacy leakage~\cite{Kuzu2011,Niedermeyer2014,Christen2017,Vidanage2019,Vidanage2020}. To preserve privacy, many approaches have been proposed that use secure multi-party computation (SMC) to privately compare pairs of records. While these methods support a large range of similarity metrics -- and maintain privacy -- they are inefficient at the scale of typical aid programs.
Differential privacy techniques can reduce the number of comparisons~\cite{He2017}, but  cannot guarantee the absence of leakage of recipients' data.
Finally, existing \ac{fpsi} protocols \cite{Uzun2021,Kulshrestha2021,Chakraborti2023,Chongchitmate2024,Gao2024,Blass2025,Bui2025,Baarsen2024,Richardson2024,Baarsen2025} solve a related problem, but require embedding registration records into a metric space. However, existing embeddings that preserve the similarity of biographical data \cite{Scannapieco2007,Li2006,Bonomi2012} typically embed into a high-dimensional Euclidean space, and existing \ac{fpsi} protocols for Euclidean space do not scale well to high dimensions.

In this paper, we propose \name, a new deduplication system that combines an embedding into Hamming space with an \ac{fpsi} protocol for Hamming space: Organizations locally transform their records into representations in Hamming space and then use an \ac{FPSI} protocol to find pairs of similar records in Hamming space. Existing \ac{FPSI} protocols are not applicable since they rely on potentially unmet assumptions on the structure of input data, approximate with insufficient accuracy, or are inefficient in our scenario (see \S\ref{sec:eval:rw}). We thus present \otfpsi, a practically efficient \ac{fpsi} mechanism that builds on SHADE~\cite{Bringer2013} and relies only on \ac{ot}. We evaluate \otfpsi extensively and show that, in this setting, \otfpsi outperforms all proposed \ac{fpsi} protocols. The main strength of \otfpsi is not only that it is more efficient for many parameters, but it does so while returning exact results and without assumptions on the structure of input data.  

For our target size, our system takes \qty{3}{\hour} to perform deduplication. This is a reduction by \qty{84}{\x} compared to existing methods (see \S\ref{subsec:eval:pprl}).   
For ethical reasons, we evaluated our system on a synthetic dataset and did not work with real humanitarian data. We modeled duplicates based on common errors and show that \textsf{xDup}'s embedding with an exact FPSI protocol only misses \qty[round-precision=1]{0.574}{\percent} of duplicates. 

\descr{Our Contribution.} We summarize our contributions.
\begin{itemize}[nosep, wide=0.5\parindent]
\item[\checkmark] We gather and formalize requirements for a humanitarian deduplication system working on biographical data based on literature and conversations with NGOs (\S\ref{sec:system_model}).

\item[\checkmark] We propose \name, an end-to-end private deduplication system that is fit for use by humanitarian organizations (\S\ref{sec:system}). 

\item[\checkmark] We show that deduplication can be reduced to Hamming \ac{fpsi}, retaining the accuracy of plaintext matching (\S\ref{sec:embedding})

\item[\checkmark] We present \otfpsi, an OT-based \ac{fpsi} protocol that does not rely on input assumptions (\S\ref{sec:ot_fpsi}). 

\item[\checkmark] Our extensive benchmarks show that this approach is more efficient than all existing FPSI protocols (\S\ref{sec:eval:rw}).

\item[\checkmark] We evaluate the end-to-end cost of \name and show that it satisfies the real-world humanitarian requirements (\S\ref{subsec:eval:e2e}).
\end{itemize}

\section{System Overview}
\label{sec:system_model}
We present the system model and design overview of \name. The definition of problem, entities, and  requirements result from a review of humanitarian deduplication~\cite{haffar2022, haffar2022b, haffar2023, haffar2023b, CALP, DIGID, DIGIDslides, Douglas2023, Douglas2023b,WFPBB, WFPBreport, OCHA2022, ifrc2023} and several discussions with humanitarian organizations.

\subsection{Entities}\label{sec:syst:entities}
Our deduplication system involves the following entities:

\descr{Field Teams.} Field teams are responsible for providing humanitarian aid (e.g., food, essential items, services) to aid recipients. Field teams can be regional and country offices of large international humanitarian organizations (e.g., ICRC, UN OCHA, MSF, and UNRWA), or local organizations (e.g., national societies of the Federation of the Red Cross). Field teams operate aid programs, and register recipients to whom they provide aid. Multiple independent field teams can operate in the same area.
Field teams are local, often operating in difficult circumstances in crisis-affected areas with limited digital resources: Hardware might be limited to laptops or simple desktops, and internet connectivity may not be reliable. To effectively distribute aid, field teams typically rely on access to their recipients' registration data. 

\descr{Headquarters.} Many field teams are part of a larger international humanitarian organization (NGO), whose headquarters (e.g., located in Geneva, Switzerland, or New York) have access to better resources and connectivity. Headquarters do not directly take part in the aid distribution or deduplication process. Yet, they want to ensure a fair distribution of humanitarian aid.
We use headquarters to provide the computing resources and connectivity necessary to operate our privacy-friendly deduplication system. Headquarters of large organizations may be protected by privileges and immunities~\cite{leblond2018immunity}.

\descr{Recipients.} People in crisis-affected areas want to receive aid from humanitarian organizations. To do so, they register with a field team or aid program as an aid recipient. As part of this process, they provide basic biographical information (e.g., first and last name, date of birth, place of origin, and information about the household composition). Field teams use this information to register recipients and allocate and distribute appropriate assistance.
As a result of the field conditions, the recorded biographical information often contains errors. For example, names might be recorded with slight variations due to differences in transcribing, and dates of birth are sometimes approximated because the true date of birth is unknown. As registration is a manual process, simple typos can also occur. Deduplication should work despite such differences in records. We assume that registration data is not maliciously incorrect (see \S\ref{subsec:system:limitations}).

Additionally, strong unique identifiers like personal ID numbers or phone numbers are often not available, or unreliable. While recipients are more likely to have a phone number than a personal ID, these numbers are subject to frequent change or shared, especially as people move around. When available, field teams record these identifiers, but this is often not the case. Therefore, in our work, \emph{we assume unique identifiers are not available}.

\begin{figure}[t]
    \centering
    \includegraphics[scale=0.2]{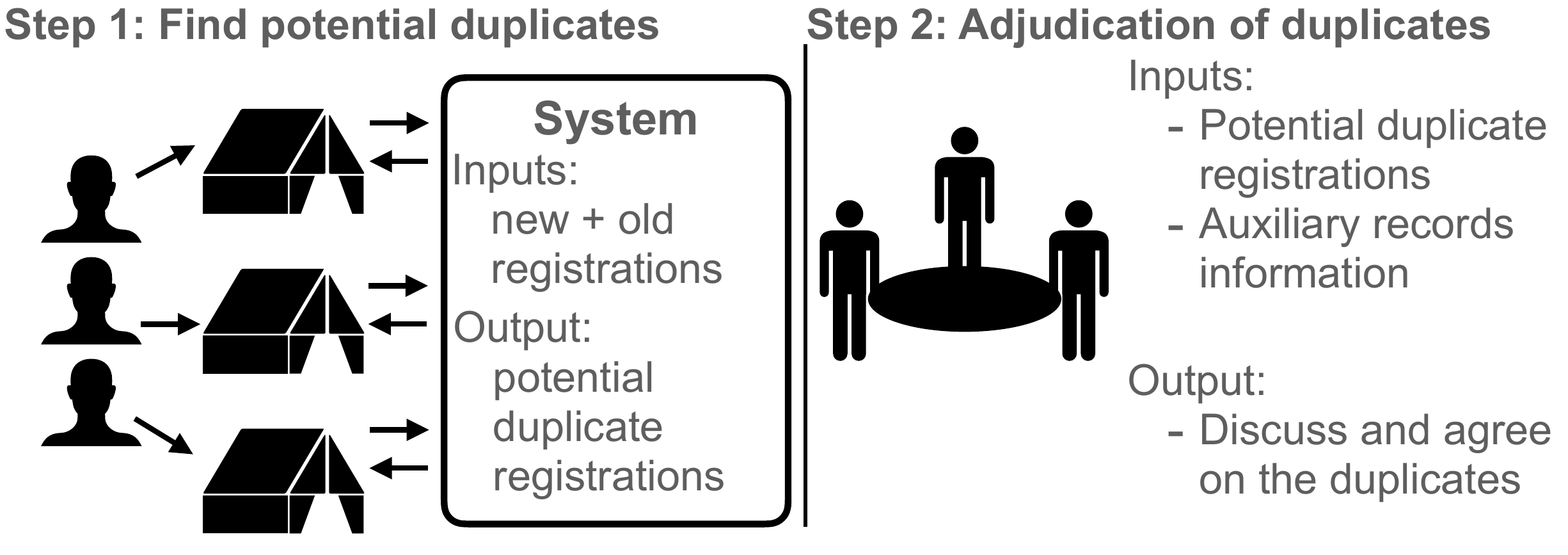}
    \caption{{High-level deduplication process}}
    \label{fig:dedup}
\end{figure}

\subsection{Overview of Humanitarian Deduplication}\label{sec:syst:workflow}

We outline the high-level registration and deduplication process resulting from conversations with NGOs and as described in documents published by NGOs~\cite{haffar2022, DIGID, Douglas2023}. 
Most organizations currently use an \emph{asynchronous} deduplication process, which \name supports. Yet, \name can also provide an online deduplication mechanism (like Janus~\cite{Edalatnejad2024}), but this still requires asynchronous manual verification.

\descr{Step 0. Registering Aid Recipients.} Field teams register aid recipients for the aid programs they operate. As part of the registration process, and to fit recipients' needs, field teams collect biographical information (names, date of birth, etc.) from aid recipients. As explained above, this information is not necessarily fully correct, and small errors are possible.
During the registration process, field teams immediately perform \emph{local deduplication} to verify that the new recipient did not already register with them.

\descr{Step 1. Identifying Potential Duplicates.}
Because NGOs have limited resources to provide assistance, they wish to help as many people as possible. Thus, they want to detect recipients that register -- purposefully or not -- with multiple teams and would unfairly receive additional assistance.

The goal of our system is to identify these \emph{cross-organizational} duplicates, i.e., newly registered recipients that are also registered with any other field team active in the same region. Because registration data can be inconsistent, the deduplication process must be robust to small differences in registration data.
It is this \emph{identification of potential cross-organizational duplicates} that we focus on in our work. As we explain in Section~\ref{sec:related_work}, current approaches fail to protect the privacy of recipients, are impractical, or fail to detect (most) duplicates. \name provides strong privacy protection, is efficient, and finds 
\qty{099.426}{\percent} of duplicates.

As field teams may not have access to reliable network connections, the system needs to support offline operation: The field teams need to be able to perform the registration offline and submit their registrations to the deduplication system at a later time.

However, if a network connection is available, an online operation mode is preferable so that the field team immediately learns about possible duplicates. This feedback allows field teams to directly gather additional information from the recipient -- which may be especially useful in cases of accidental duplicate registrations. 

\name supports both modes of operation: an offline mode to deduplicate a batch of new registrations, and an online mode to deduplicate a single new registration in real-time. 

\descr{Step 2. Verifying Duplicates.}
The final step is to verify which potential duplicates are true duplicate registrations. This is a manual process: In fixed intervals, the \emph{deduplication committee} gathers and discusses the potential duplicates \cite{DIGID} (independent of whether they were discovered in online or offline mode).
Each field team sends a representative who has access to the list of new potential duplicates as well as that team's full registration information. For each identified duplicate, the representatives compare the full registration data to assess whether this recipient is truly a duplicate. 
The manual nature of this process rules out potential false positives, ensures that field teams can incorporate all information available about aid recipients (not all of this information is necessarily used during step 1), and that appropriate measures can be taken when they do detect duplication.

\subsection{Goals and Non-Goals}\label{sec:syst:goals}

The goal in our work is to build a \emph{cross-organizational deduplication system} for humanitarian organizations that uses biographical data to determine potential duplicates.

\descr{Ideal Functionality.} We formalize the deduplication functionality we aim to provide:
A \emph{querying} organization wishes to determine which of their new registrations are potential duplicates in the set of all registrations held by a \emph{responding} organization. To this end, the querying organization inputs a single new registration (in online mode) or a \emph{batch} of new registration records (in offline mode), and the responding organization inputs \emph{all} registration records (new and old). Our functionality compares records and outputs which querier record's similarity to a responder record exceeds a threshold.

\descr{Non-goals.} From discussions with NGOs and analysis of their requirements, we made the following design decisions.

\noindent\emph{Not an automated decision-making system.} We deliberately did not design an automated decision-making system. Our goal is only to identify potential duplicate registrations, that subsequently have to be manually checked in an \emph{adjudication process}~\cite{DIGIDslides, DIGID, ifrc2023}.

\noindent\emph{Do not rely on unique identifiers.} Our system has been designed to function in a common setting where reliable unique identifiers (e.g., personal ID or phone numbers) are unavailable. When such identifiers are available~\cite{Wille2023}, simpler solutions are possible.

\subsection{Requirements}\label{sec:syst:reqs}

We summarize functional, security, privacy, and deployment requirements for cross-organizational deduplication identified from humanitarian publications and discussions with humanitarian organizations.

\descr{Functional Requirements.} \name must satisfy the following:

\reqdef{F1}{recall}{Identification of Duplicates} The system should identify which of the newly registered recipients of one field team are also registered with another field team. It should do so with high recall.

\reqdef{F2}{noid}{No IDs} The system should not rely on unique fixed identifiers for the recipients. 

\reqdef{F3}{fuzzy}{Fuzzy matching} The system should support fuzzy matching on quasi-identifers (e.g., name, DoB, gender).

\descr{Privacy Requirements.} To protect the privacy of vulnerable recipients, \name must provide the following properties.

\reqdef{P1}{fpr}{Low False-Positive Rate} 
The system should have a low false-positive rate (FPR), i.e., ensure that very few of the new registrations are falsely flagged as duplicates.
A low FPR reduces the privacy impact on non-duplicate recipients. Recall that, for each potential duplicate identified in step 1, the organization subsequently shares this data with other organizations in step 2. The fewer duplicates our system incorrectly identifies, the better we can protect privacy.
A low FPR also reduces the workload on the deduplication committee. 
We thus aim for an FPR of \qty{0.1}{\percent} to ensure that only a small fraction of the discussed potential duplicates turns out to be false.

\reqdef{P2}{leakage}{No Leakage} During deduplication, the responding field team should learn no information about the queried records and the querying field team should learn nothing about non-matching responder records. The headquarters should learn no information about individual registrations.

\descr{Deployment Requirements.}
We require our system to be suitable for real-world deployment. 

\reqdef{D1}{offline}{Support Offline Operation} Field teams operate in challenging environments in which internet access may be unreliable. Thus, any system should support an offline mode where field teams submit a batch of queries and later retrieve responses, without requiring them to be online.

\reqdef{D2}{online}{Support Online Operation} If field teams have network access during registration, the system should support online operation, performing deduplication of a single record within seconds; thus enabling the field team to take immediate action (such as requesting more information).

\reqdef{D3}{local-efficient}{Efficient for Field Teams} The system should work with the limited compute and communication resources available to field teams.

\reqdef{D4}{scalability}{Scalability} The system should be able to cope with realistic population sizes. A single humanitarian program typically serves less than $100$k people~\cite{Edalatnejad2024, WangLSGT23, WFPBB}, and we assume that submitted batches in offline mode contain up to around $2$k new registrations.

\descr{Current Deduplication Does not Satisfy these Requirements.}
The approaches used by humanitarian organizations right now (if any) for cross-organizational deduplication do not satisfy the requirements set out above. Methods based on direct data sharing or plaintext similarity matching fail to satisfy the privacy requirement \reqlinky{leakage} because they potentially reveal a lot of registration information about non-duplicates.
To reduce leakage, some humanitarian actors instead apply cryptographic hash functions to all (or a carefully chosen subset) of the registration data and then share these hashes~\cite{haffar2023, Douglas2023b, ifrc2023}. While this is better than directly sharing the data, these hashes are still vulnerable to membership inference attacks (where it is trivial to check whether a specific person appears) as well as brute-force reconstruction attacks. As a result, these approaches do not satisfy \reqlinky{leakage}. Moreover, as a result of applying a hash-function, small changes in the records can now result in duplicates not being found. Thus, these approaches cannot provide high recall (violating \reqlinky{recall}) or can do so only at the cost of many false positives (violating \reqlinky{fpr}).

\subsection{Threat Model}\label{sec:syst:threat}

\descr{Headquarters. } 
Large NGOs like the UN or ICRC are protected by privileges and immunities~\cite{leblond2018immunity}. While we assume that their headquarters are resistant to coercion, they may still be compromised~\cite{icrc2022attack, Guardian2022}. We model headquarters as honest-but-curious and assume the organizations' headquarters do not collude with each other.

\descr{Field Teams. } We assume that field teams perform the recipient registration honestly since biographical deduplication relies on the trustworthiness of registration data. Yet -- because field teams operate in challenging circumstances and are therefore vulnerable to compromise and coercion (e.g., by local actors) -- we consider them malicious in the deduplication process to ensure that coercion of one field team does not reveal information about recipients registered with \emph{other} field teams.  

\descr{Recipients. } Similar to the NGOs' current processes \cite{DIGID}, we assume that there is a verification mechanism in place to ensure the validity of registration data, and thus most errors are accidental. To ensure validity, humanitarian organizations often consult appropriate sources -- for example, elders in the communities that these organizations target. 

\subsection{Limitations}
\label{subsec:system:limitations}

Any deduplication system brings privacy risks through the ideal deduplication functionality. We acknowledge these risks and stress that they are inherent to all deduplication systems and must be mitigated using out-of-band measures.

\subdescr{Malicious Registrants. } Deduplication based on biographical data hinges on self-reported recipient data being trustworthy and, hence, organizations need a mechanism to enforce correctness. 
If registration data cannot be trusted, e.g., because recipients can lie without being detected, deduplication methods based on biographical data are inappropriate. In practice, organizations have found such validation mechanisms \cite{DIGID,ifrc2023} and use biographical data for deduplication.

Additionally, malicious registrants could abuse the deduplication system to extract information about other individuals: They could try to register with another individual's personal data to find out whether this individual is already registered with any organization. This attack can only be avoided if there are mechanisms in place to ensure registrants cannot lie during registration.

\subdescr{Compromised Field Teams. } Every query inherently reveals some information about the responder's database. A compromised field team may, e.g., perform a dictionary attack to enumerate the databases of other organizations. Every deduplication mechanism is vulnerable to such attacks, and their impact can only be controlled through rate imitating.

\begin{figure}[t]
    \centering
    \includegraphics[width=0.8\columnwidth]{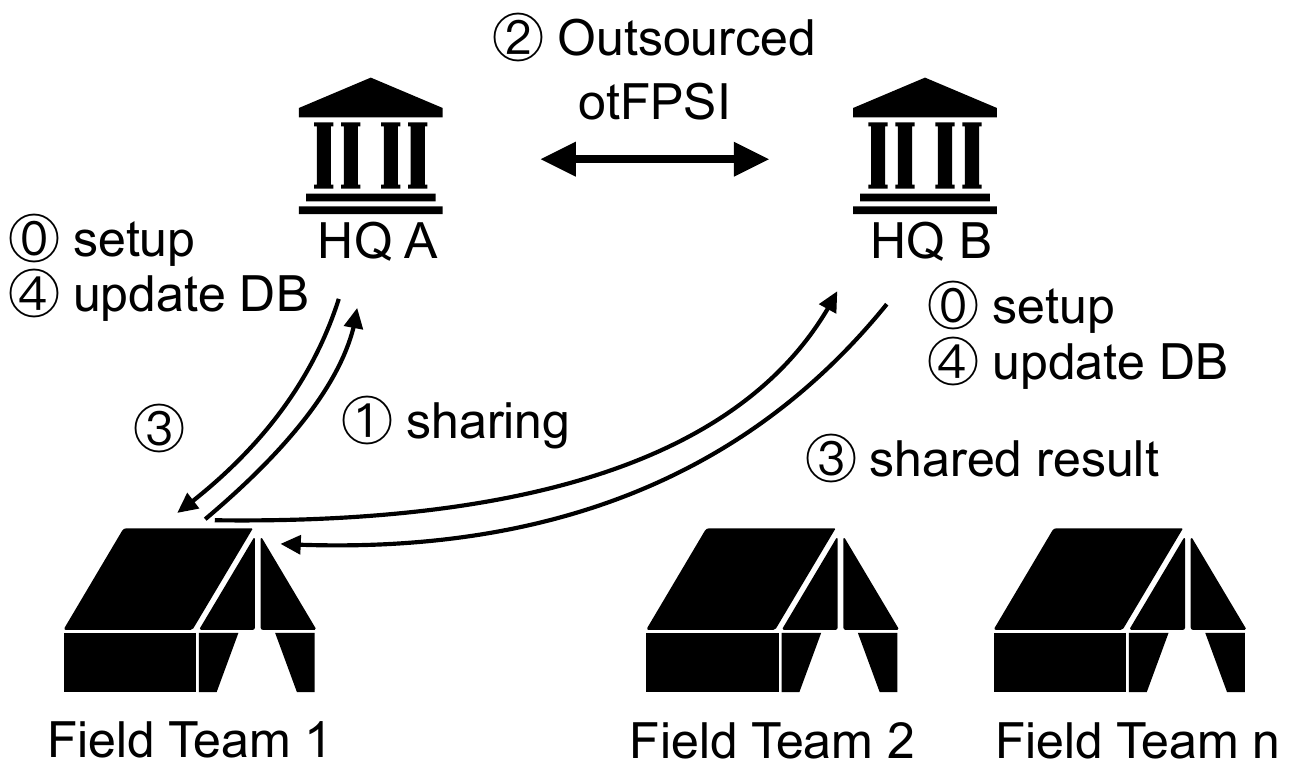}
    \caption{Illustration of \name.}
    \label{fig:xDup}
\end{figure}

\subsection{Design Overview}\label{sec:syst:design}
We address the requirements set out in the previous section: To maintain privacy (\reqlinky{leakage}) we use a cryptographically-secure matching mechanism to compare individual records. However, existing matching protocols using generic \ac{SMC} or \ac{he} are too costly to fulfill the scalability requirement (\reqlinky{scalability}). This is especially the case for the online operation mode, where the responding party inherently needs to perform computation linear in the database size. 
Many existing mechanisms to reduce the number of comparisons typically assume that the querier holds a set of records instead of just one, and are not applicable in our online mode.

To overcome these limitations, we first transform the structured registration records into fixed-length bit strings, such that similar records have a small Hamming distance. We then use our new \otfpsi protocol, an \ac{fpsi} protocol for Hamming space, to privately compare the embedded records. \otfpsi utilizes a concretely efficient matching mechanism built on \ac{ot}.

To address the challenge of limited computational resources (\reqlinky{local-efficient}) and online/offline requirements (\reqlinky{offline} and \reqlinky{online}), see Figure~\ref{fig:xDup}, we outsource the computation to two more powerful compute nodes operated by two organizations' headquarters, each holding secret-shared databases of the embeddings of all field teams' registration records \Circled{0}. When a field team wants to use \name to check one or multiple new registrations, it locally computes their embedding and sends secret shares to both compute nodes \Circled{1}. The compute nodes then run an outsourced variant of \otfpsi to compare the new registrations to all registrations in their databases \Circled{2}. Finally, they send the secret-shared result back to the querying field team \Circled{3} and add the new registrations to their databases \Circled{4}.

\section{Related Work}
\label{sec:related_work}

As mentioned in \S\ref{sec:syst:reqs}, current deduplication mechanisms for NGOs rely on collision-resistant hash functions. Yet, this approach (i)~leaks personal information about the recipients, and (ii)~is not robust to slight perturbations in the attributes that can naturally occur during registration.

\subsection{\acl{pprl}}
\acreset{pprl}

To solve the privacy issue, NGOs could rely on \ac{pprl}: In this setting, two (or more) parties hold databases of records and want to identify all pairs of records -- with potentially varying sets of attributes in different databases -- that correspond to the same individual~\cite{Christen2012}. Most \ac{pprl} approaches rely on a matching functionality that compares two records.

To ensure robustness to small perturbations of attributes, early works use different similarity metrics built on top of Bloom filters~\cite{Lai2006,Schnell2009,Durham2010,Karakasidis2011,Schnell2011,Vatsalan2012b,Durham2014,Ranbaduge2014,Vatsalan2014}. Yet, revealing these Bloom Filters to other parties without additional privacy mechanisms is vulnerable to attacks~\cite{Kuzu2011,Niedermeyer2014,Christen2017,Vidanage2019,Vidanage2020}.

A different research direction provides private implementations of matching using homomorphic encryption~\cite{Inan2008,Kuzu2013,kasyap2024}; and generic \ac{SMC} techniques~\cite{Lazrig2018,Stammler2020,chen2018}, or PSI~\cite{FreedmanNP04,PinkasSZ18,YeSPW09}. Yet, comparing all pairs of records of two datasets using these relatively expensive matching protocols is costly and impractical for our scenario: \Mainzelliste's \ac{SMC}~\cite{Stammler2020} would require about 10 days to perform the same task that our construction can do in hours (see \S\ref{subsec:eval:pprl}).

To reduce the number of potentially costly comparisons, several works use a \emph{blocking} mechanism that identifies candidate pairs and then only apply matching to these candidate pairs~\cite{Karapiperis2015,Scannapieco2007,He2017}. Yet, this can lead to leakage about non-matching records~\cite{Cao2015} and cannot always guarantee that all matching pairs are identified, leading to false negatives. 
A popular way to implement blocking is by deterministically assigning records to buckets and only comparing records assigned to the same bucket. Yet, the composition of these buckets can reveal information. This issue is typically addressed using differential privacy and variants thereof \cite{Inan2012,He2017} but without strong cryptographic privacy guarantees.

Wei and Kerschbaum~\cite{Wei2023} present a blocking mechanism that provides cryptographic security. It uses bucketization with \emph{frequency smoothing}~\cite{Grubbs2020} in combination with \emph{private bin join}~\cite{Krastnikov2020}.
Still, their approach leaks some information via the number of performed comparisons. 
While offering good performance, their implementation currently does not perform any fuzzy matching (i.e, it only considers strict equality of 16-bit integers). Thus, it is unclear how this solution would perform in real-world record linkage use cases involving larger data sizes and fuzzy matching. 

Finally, blocking mechanisms typically compare two sets of records -- which only applies to our offline operation mode. For online operations with only a single query, blocking mechanisms do not improve performance.

In a different vein, \ac{lsh} can reduce \ac{pprl} to \ac{psi} \cite{Adir2022,Han2025}. 
    We evaluate this approach and observe that it does not provide the required accuracy in our setting -- it provides only \qty{086,46240234375}{\percent} recall compared to our %
    \qty{099.426}{\percent} (see \cref{ap:jacard_lsh}).

More works on \ac{pprl} exist, yet many do not provide strong security guarantees or are prohibitively expensive. We refer readers to surveys for details~\cite{Gkoulalas-Divanis2021, Vatsalan2013}.

\subsection{\acl{fpsi}}
\label{subsec:rw:fpsi}
\acreset{fpsi}

Instead of \ac{pprl} techniques, NGOs could also rely on modern \ac{FPSI} approaches. While \ac{psi} computes the intersection of two sets, \ac{fpsi} computes which elements are close with regard to a distance metric \distance and a threshold \threshold.
In our setting, the parties would individually transform their records to a metric space (e.g., Euclidean or Hamming space) such that matching records are close in that metric space. Then, the parties use a (compatible) \ac{fpsi} protocol to find matches while preserving the privacy of non-matching records.

Several works exist that transform records into Euclidean space \cite{Scannapieco2007,Li2006,Bonomi2012}. However, these approaches result in high-dimensional embeddings -- for our NGOs' setting, we expect a dimensionality of more than 50 (see \cref{ap:rw_euclidean}). 

The embeddings into Euclidean space could be combined with an \ac{fpsi} protocol for Euclidean space \cite{Baarsen2024,Gao2024,Richardson2024,Baarsen2025}. Yet, these protocols come with significant drawbacks: They place potentially restrictive assumptions on the structure of the input data and many of these protocols do not scale well to high dimensions.
For instance, the state-of-the-art work by Van Baarsen and Pu~\cite{Baarsen2025} proposes two protocols. The first requires the parties' data points to be at least $2\threshold\sqrt{\dimension}$ or $2\threshold(\sqrt{\dimension} + 1)$ apart, but has a runtime linear in $2^\dimension\dimension$, where \dimension is the data dimension and \threshold the distance threshold. We infer from their work that the cost of this protocol is prohibitively high for $\dimension \ge 50$. 
Their second protocol, which is linear in $\dimension\threshold$, and thus has better asymptotics, relies on the even stronger assumption that each data point's projections on each dimension are at least $2\threshold$ apart from all other points. We cannot rely on this assumption to hold for large datasets with existing embeddings. Similar limitations also apply to other FPSI protocols for Euclidean space \cite{Baarsen2024,Gao2024,Richardson2024}.

Another line of \ac{fpsi} protocols \cite{Uzun2021,Kulshrestha2021,Chakraborti2023,Chongchitmate2024,Gao2024,Blass2025,Bui2025} operates in Hamming space. 
However, these FPSI protocols have significant drawbacks: Some approximate the Hamming distance and do not achieve the accuracy required in our setting \cite{Uzun2021,Chakraborti2023,Bui2025}. For our embedding, we need a relatively high-dimensional Hamming space ($\dimension \approx 512$) and a high distance threshold ($\threshold \approx \dimension/4$) (see \cref{ap:emebdding}). For these parameters, existing \ac{fpsi} protocols have unfulfillable input assumptions \cite{Gao2024,Chongchitmate2024}, or are inefficient since their runtime depends on the threshold or is super-linear in the dimension \cite{Uzun2021,Chakraborti2023,Gao2024,Blass2025,Bui2025} (see \S\ref{sec:eval:rw}). 
While using an embedding with an \ac{fpsi} protocol seems a promising direction, existing works can not be easily combined.

\section{\name}
\label{sec:system}
We now present the design overview of \name. We present the high-level building design rationale, introduce our building blocks, and detail our system design.

\subsection{Design Rationale}
\label{subsec:system:design}

\begin{figure}
    \centering
    \fbox{
        \begin{minipage}{0.9\columnwidth}
            \footnotesize
            \textbf{Parameters:}  Dimension \dimension, distance metric \distance, threshold \threshold, set sizes \sizeA{} and \sizeB
            \begin{enumerate}[leftmargin=*]
                \item Receive $\setA = \{\elementA_1, \dots, \elementA_\sizeA\} \subseteq \bin^\dimension$ from \partyA and $\setB = \{\elementB_1, \dots, \elementB_\sizeB\} \subseteq \bin^\dimension$ from \partyB.
                \item Send $\{(i, j) \mid i \in \range{\sizeA}, j \in \range{\sizeB}, \distance(\elementA_i, \elementB_j) \le \threshold\}$ to \partyA.
            \end{enumerate}
        \end{minipage}
    }
    \caption{\ffpsi, Ideal functionality for FPSI between querier \partyA with input \setA and responder \partyB with input \setB. $\range{n}=\{1, ..., n\}$.}
    \label{fig:sytem:fpsi_ideal}
\end{figure}

One of the design challenges of \name is to provide a query mechanism that allows one organization to perform a query when all other organizations may be offline (\reqlinky{offline}). This requirement and field teams' limited resources (\reqlinky{local-efficient}) preclude the direct use of interactive \ac{smc} protocols. 

While \ac{HE} appears to be auspicious for this model -- as it might allow outsourcing to a single untrusted server -- it also brings challenges.
First, the key management is non-trivial: under which key are the ciphertexts encrypted, who performs the decryption, etc. 
Second, secret-key holders must be online for decryption. One potential solution would be to operate under the querying organization's key. To guarantee privacy in this setting, the querying organization must not collude with the compute server. Yet, as the compute server will likely be operated by one of the organizations, this non-collusion assumption may be hard to warrant. 

A non-collusion assumption between two servers operated by two different organizations is a more natural fit for the humanitarian setting. These servers may be operated by the organizations' headquarters, which typically have sufficient resources available, want to assist the aid distribution process, and, for some organizations, are protected (e.g., against coercion) by special privileges and immunities~\cite{leblond2018immunity}.

Thus, \name relies on outsourcing the computation and communication of its interactive FPSI protocol to two non-colluding compute nodes operated by two headquarters. 

This design has another advantage: It remains secure if field teams act maliciously -- all they can do is send queries to the compute nodes (which may still leak, see \S\ref{subsec:system:limitations}). 

\subsection{Building Blocks}
\label{subsec:system:blocks}
 
We rely on a novel approach that combines an embedding mechanism into Hamming space with an \ac{fpsi} protocol. This approach enables us to provide high performance (using efficient \ac{fpsi} protocols), while being agnostic to the properties of the records (using a suitable embedding).

\descr{Embedding.} Given a universe of records \recorduniverse, an embedding $\embedding: \recorduniverse \rightarrow \bin^\dimension$ maps records to fixed-length bit strings. An embedding should map two records $r, r' \in \recorduniverse$ that match (i.e., correspond to the same individual) to similar bit strings. This means that $\hammingdist(\embedding(r), \embedding(r')) \le \threshold$ where \hammingdist denotes the Hamming distance and \threshold is a constant threshold.

\descr{Fuzzy Private Set Intersection.} 
\Cref{fig:sytem:fpsi_ideal} formalizes $\mathcal{F}_{\text{FPSI}}$, the ideal functionality of FPSI for identifying which elements from \partyA and \partyB are close w.r.t. a distance metric \distance and a threshold \threshold. To allow outsourcing computation to two untrusted compute nodes, \name uses an \ac{fpsi} protocol that can operate on secret-shared inputs and outputs. We formalize this functionality in \cref{fig:sytem:ssfpsi_ideal}. In secret-shared FPSI, two compute nodes \serverA and \serverB each hold one secret share of each of the input sets \setA and \setB. Secret-shared FPSI first reconstructs these shares, compares all records, and finally outputs one share of the result to each party.

\begin{figure}
    \centering
    \fbox{
        \begin{minipage}{0.9\columnwidth}
            \footnotesize
            \textbf{Parameters:}  Dimension \dimension, distance metric \distance, threshold \threshold, set sizes \sizeA{} and \sizeB
            \begin{enumerate}[leftmargin=*]
                \item Receive $\shareA{\setA} = \{\shareA{\elementA_1}, \dots, \shareA{\elementA_\sizeA}\}, \shareA{\setB} = \{\shareA{\elementB_1}, \dots, \shareA{\elementB_\sizeB}\} \subseteq \bin^\dimension$ from \serverA and $\shareB{\setA} = \{\shareB{\elementA_1}, \dots, \shareB{\elementA_\sizeA}\}, \shareB{\setB} = \{\shareB{\elementB_1}, \dots, \shareB{\elementB_\sizeB}\} \subseteq \bin^\dimension$ from \serverB.
                \item Compute $\elementA_i = \shareA{\elementA_i} \xor \shareB{\elementA_i}$ and $\elementB_j = \shareA{\elementB_j} \xor \shareB{\elementB_j}$ for $i \in \range{\sizeA}, j \in \range{\sizeB}$.
                \item Sample $\shareB{M} \sample \bin^{\sizeA \times \sizeB}$ and send $\shareB{M}$ to \serverB.
                \item Compute $\shareA{M}$ as $\shareA{M}_{i,j} = \shareB{M}_{i,j} \xor (\distance(\elementA_i, \elementB_j) \le \threshold)$ for $i \in \range{\sizeA}, j \in \range{\sizeB}$ and send $\shareA{M}$ to \serverA.
            \end{enumerate}
        \end{minipage}
    }
    \caption{\fssfpsi, Ideal secret-shared FPSI functionality between node \serverA with input $\shareA{\setA}, \shareA{\setB}$ and node \serverB with input $\shareB{\setA},\shareB{\setB}$.  The output is secret-shared across $\shareA{M}$ and $\shareB{M}$.}
    \label{fig:sytem:ssfpsi_ideal}
\end{figure}

\subsection{System Description}

We now present \name in more detail. We assume there are \norgs field teams $\organization{1}, \dots, \organization{\norgs}$. To enable online queries, the compute nodes \server{1}, \server{2} hold a secret-shared database of all registrations that is continuously updated after each query.

\descr{Parameter Selection.} All field teams agree on the following parameters of the system:
\begin{itemize}[leftmargin=*, topsep=0pt] 
    \item \emph{Embedding}: An embedding mechanism to transform records to Hamming space with dimension \dimension.
    \item \emph{Compute Nodes}: Two non-colluding nodes \server{1} and \server{2}. This role can be taken by two headquarters (see \S\ref{sec:syst:entities}). 
\end{itemize}

\descr{One-Time Setup.} In the setup phase (\cref{fig:system:code_setup}) each field team \organization{i} submits its pre-existing registration database  (which is assumed to be without duplicates) to the compute nodes. To do so, \organization{i} embeds all records in its registration database \registrationdb{i} into Hamming space, creates secret shares of the embeddings, and sends them to the compute nodes \serverA and \serverB. 

\begin{figure}
    \begin{pchstack}[center,space=10pt,boxed]
    \procedure[codesize=\footnotesize,headheight=2.5ex]{\small \organization{i}: One-Time Setup}{
        L_1, L_2 \gets [\,] \\
        \pcfor r \in \registrationdb{i}: \\
            \t s \sample \bin^\dimension\\
            \t L_1\mathsf{.append}(s) \\
            \t L_2\mathsf{.append}(s \xor \embedding(r)) \\
        \text{Send $L_1$ to \server{1} and $L_2$ to \server{2}}
    }
    \procedure[codesize=\footnotesize,headheight=2.5ex]{\small \server{i}: One-Time Setup}{
        \pcfor j \in \range{\norgs}: \\
            \t \text{Receive $D_j$ from \organization{j}} \\
    }
    \end{pchstack}
    \caption{One-time setup procedures with embedding \embedding.}
    \label{fig:system:code_setup}
\end{figure}

\descr{Deduplication.} To query \name with a set of new registrations $Q$ (which contains only one element in the online case), field team \organization{i} performs the following steps (see \cref{fig:system:code_deduplication}):

\begin{enumerate}[leftmargin=*, topsep=0pt]
    \item \emph{Local Deduplication}: First, \organization{i} locally deduplicates, that is, it identifies new registrations in $Q$ that are already registered with \organization{i} itself. This process happens locally and, hence, is done in plaintext and may happen each time a new registration is recorded.
    \item \emph{Query}: \organization{i} embeds its query set $Q$, creates secret shares, and sends one each to \server{1} and \server{2}.
    \item \emph{Process}: \server{1} and \server{2} run a secret-shared FPSI protocol to compare \organization{i}'s new registrations $Q$ to all stored registrations of the other field teams \organization{j \neq i}. Both compute nodes append the secret shares of the new registrations to the existing registrations of the querying field team.
    \item \emph{Retrieve}: \organization{i} retrieves the secret-shared results from \server{1} and \server{2} and reconstructs the result. For each query record, it identifies whether there is a duplicate and, if so, with which other field team. 
\end{enumerate}

\begin{figure}
    \centering
    \begin{pcvstack}[boxed]
        \begin{pchstack}[center,space=10pt]
            \procedure[codesize=\footnotesize,headheight=2.5ex]{\small \organization{i}: Query}{
                Q_1, Q_2 \gets [\,] \\
                \pcfor q \in Q: \\
                    \t s \sample \bin^\dimension\\
                    \t Q_1\mathsf{.append}(s) \\
                    \t Q_2\mathsf{.append}(s \xor \embedding(q)) \\
                \text{Send $Q_1$ to \server{1} and $Q_2$ to \server{2}}
            }
            \procedure[codesize=\footnotesize,headheight=2.5ex]{\small \server{i}: Process}{
                \text{Receive $Q_i$ from \organization{j}} \\
                D_{\neq j} \gets [\,] \pccomment{Collect registrations} \\
                \pcfor k \in \range{\norgs} \setminus \{j\}: \\
                \t  D_{\neq j}\mathsf{.insert}(D_k) \\
                M^{(i)} \gets \mathsf{ssFPSI}_i(D_{\neq j}, Q_i) \\
                L \gets [D_k\mathsf{.length}() \mid k \in \range{\norgs}] \\
                L[j] \gets 0 \pccomment[1em]{maps records to orgs}\\
                D_j\mathsf{.insert}(Q_i) \pccomment{save new data} \\
                \text{Make $M^{(i)}$, $L$ available to \organization{j}}
            }
        \end{pchstack}
        \begin{pchstack}[center]
            \procedure[codesize=\footnotesize,headheight=2.5ex]{\small \organization{i}: Retrieve}{
                \text{Retrieve $M^{(1)}$ and $L$ from \server{1}, and $M^{(2)}$ from \server{2}} \\
                n_Q,n_R \gets M^{(1)}\mathsf{.size()} \\
                \pcfor q \in \range{n_Q}: \\
                    \t o \gets 1; \pccomment{organization counter} \\
                    \t \pcfor j \in \range{n_B}: \\
                    \t \t \pcif j > \Sigma_{k=1}^{o} L[k]: \\
                    \t \t \t o \gets o +1 \\ 
                    \t \t \pcif M^{(1)}_{q,j} \xor M^{(2)}_{q,j} = 1: \\
                    \t \t \t r = j - \Sigma_{k=1}^{o-1} L[k] \\
                    \t \t \t \mathbf{output} \text{ Duplicate query $q$ with record $r$ of \organization{o}}
            }
        \end{pchstack}
    \end{pcvstack}
    \caption{Deduplication procedures.}
    \label{fig:system:code_deduplication}
\end{figure}

\descr{Manual Verification.}  At fixed intervals, all field teams join the deduplication committee with additional information about the potential duplicates discovered in the previous step to perform the adjudication process (\S\ref{sec:syst:workflow}).

\section{Embedding}
\label{sec:embedding}

\acreset{lsh}
We are not aware of an existing embedding that matches the requirements imposed by our humanitarian use case. Hence, we use a new embedding strategy which, at its core, uses \ac{lsh}: $\dimension$ \ac{lsh} functions each convert the \qgrams (i.e., all substrings of length \gramlen) of the record into a single bit. The final embedding is the concatenation of these $\dimension$ bit digests. 
By properties of \ac{lsh}, the more similar the input records are, the more individual bit digests match. 
To compare two embeddings, we use the Hamming distance and two records are deemed duplicates if the Hamming distance is below a threshold $\threshold$. We provide more details in Appendix~\ref{ap:emebdding}.

To validate our embedding, we evaluate it using a synthetic dataset representative of humanitarian deduplication tasks (see \cref{ap:dataset}). 
The deduplication of a single record in a large database (\num[round-mode=none,exponent-mode=input]{131072} records) leads to a false-negative rate of \qty{0.57}{\percent} at a false-positive rate of \qty{0.098}{\percent} (\reqlinky{fpr}) (with embedding size of $\dimension = 511$ bits and a Hamming distance threshold of $\threshold = 132$).
We consider these accuracy results to fulfill the \name's requirements (\reqlinky{fpr}, \reqlinky{recall}) and choose $\dimension = 511$ and $\threshold = 132$ as the operating parameters for \name. For lower dimensions, we can only achieve higher false-negative rates at the target false-positive rate, but still need $\threshold \approx \dimension/4$.
The accuracy of our embedding is on par with existing plaintext matching algorithms (\cref{ap:emebdding}). Nevertheless, \name is agnostic to the concrete embedding used, and this construction may be easily replaced.

\section{\otfpsi}
\label{sec:ot_fpsi}

Since existing FPSI protocols are not suitable for our purpose (\S\ref{subsec:rw:fpsi}), we introduce \otfpsi, a Hamming \ac{fpsi} protocol that is correct, assumption-free, threshold-independent, and quasi-linear in the dimension. At its core, our protocol utilizes the SHADE construction~\cite{Bringer2013} construction to privately compute Hamming distances. We combine SHADE with an efficient threshold comparison step, extend it to support secret-sharing, and enhance it with a batching method for secret-shared FPSI.
We evaluate the performance of \otfpsi in \S\ref{sec:eval:otfpsi} and show it outperforms all existing protocols.

\subsection{Oblivious Transfer}
\label{sec:fpsi:ot}
A key building block of \otfpsi is \acf{ot}. A 1-out-of-$N$ \ac{ot} is a two-party functionality between a responder \partyB holding $N$ messages $m_0,\dots,m_{N-1}$ and a querier \partyA with a choice index $c \in \ZZ_N$. \ac{ot} enables \partyA to learn $m_c$ while hiding the choice $c$ from \partyB and all other messages $m_i$ for $i \in \ZZ_N \setminus \{c\}$ from \partyA. Different \ac{ot} functionalities are classified by how much control the responder has over the messages. In chosen \ac{ot}, messages are chosen by the responder, while in random OT, they are chosen at random by the protocol. Correlated OT chooses one message at random and derives the remaining messages using correlation functions. 
    We specify the functionality of OT variants in \cref{ap:ot}.

\subsection{Protocol Description}

\otfpsi computes a secure comparison between two bit strings (i.e., the embeddings): $\elementA$ held by the querier \partyA and \elementB held by the responder \partyB. The result bit $b = (\hammingdist(\elementA,\elementB) \le \threshold)$ is only known to \partyA. This comparison is applied to all pairs of records in the sets \setA and \setB to achieve \ac{fpsi} (\cref{fig:sytem:fpsi_ideal}).

Our secure comparison protocol consists of two steps: First, it computes secret-shares of the Hamming distance $\hammingdist(\elementA, \elementB)$ and, second, compares the distance to the threshold \threshold to determine the result bit. Both steps rely on \ac{OT}. 
We detail in the following subsections how these steps are performed and how we can optimize them.

\subsection{A Single Comparison}
\label{subsec:fpsi:single_comparison}
For now, we only consider the distance computation and threshold comparison between two bit strings.

\descr{Model.} Assume the bit string  \elementA (resp. \elementB) is held by \partyA (resp. \partyB). Both \elementA and \elementB have $\dimension$ bits,  let $\elementA[i]$ be the $i$-th bit of $\elementA$. We set $\modulus = \dimension + 1$ (not necessarily prime). %
By $\range{n}$, we denote the set $\{1, \dots, n\}$. We denote assignment modulo \modulus as $\getsp$. %

\descr{Computing the Distance.} To compute secret-shares of the Hamming distance, we use the SHADE~\cite{Bringer2013} construction:  

\subdescr{Computation.} For each bit $i \in \range{\dimension}$, \partyB samples $m_i$ from \group and computes both $m_{i,0} = m_i + \elementB[i] \mod \modulus$ and  $m_{i,1} = m_i + (1 \xor \elementB[i]) \mod \modulus$. 
We then run a 1-out-of-2 chosen \ac{ot} (see \S\ref{sec:fpsi:ot}): \partyB inputs the messages $m_{i,0}$ and $m_{i,1}$, and \partyA inputs $\elementA[i]$ as the choice bit, and receives $d_i = m_{i,\elementA[i]}$. 
After looping over all $\dimension$ bits, \partyA computes $D = \sum_{i=1}^{\dimension}d_i$ and \partyB computes $M = \sum_{i=1}^{\dimension} m_i$. 

\subdescr{Correctness and Security.} By construction, $d_i = m_{i,\elementA[i]} = m_i + (\elementA[i] \xor \elementB[i]) \mod p$. With $\modulus > \hammingdist(\elementA, \elementB)$, it follows that $D - M \mod \modulus = \sum_{i=1}^{\dimension} \elementA[i] \xor \elementB[i] = \hammingdist(\elementA, \elementB)$. %
SHADE is secure in the semi-honest setting assuming the underlying OT is secure in the semi-honest setting \cite{Bringer2013,Bringer2014}.

\subdescr{Correlated OT.} Bringer \etal~\cite{Bringer2014} observe that correlated \ac{ot} is sufficient for SHADE. In correlated OT, the sender gets a single random value sampled by the protocol and inputs a correlation function determining the second value. 
Here, the protocol can sample $m_{1,0}, \dots, m_{\dimension,0}$ and then \partyB can compute $m_i = m_{i,0} - \elementB[i] \mod \modulus$ and $m_{i,1} = m_i + (1 \xor \elementB[i]) \mod \modulus$. Using correlated OT can reduce communication cost 
    compared to chosen OT (see \cref{ap:ot})
.

\descr{Comparing to \threshold.} To privately compare $\hammingdist(\elementA, \elementB) = D - M \bmod \modulus$ to the threshold \threshold, we combine SHADE with an additional 1-out-of-\dimension OT: \partyB computes the result bit of the comparison for all \modulus possible values of $D$: $v_i = (i - M \mod \modulus \le \threshold)$ for $i \in \group$ and \partyA gets the result $v_D$ by OT.

\subdescr{Correctness and Security. } By construction, \partyA learns $b = v_D = (D - M \mod \modulus \leq \threshold) = (\hammingdist(\elementA, \elementB) \leq \threshold)$.
The security guarantees of the threshold comparison follow directly from those of OT: \partyB learns no information and \partyA only learns the intended output bit.

\subsection{Full \otfpsi Construction}
With the comparison mechanism for bit strings in place, we now build the full \otfpsi protocol to implement the FPSI functionality (\cref{fig:sytem:fpsi_ideal}). 
Consider \partyA (resp. \partyB) holds the set of elements \setA (resp. \setB).

\begin{figure}
    \footnotesize
    \setcounter{linecounter}{1}
    \renewcommand{\arraystretch}{1.3}
    \setlisting{otfpsi}
    \begin{tabular}{@{}r@{}lc@{}l@{}}
    & \textbf{Querier} & & \textbf{Responder} \\
    & $\setA \! = \! \{\elementA_1, \ldots, \elementA_{\sizeA}\} \! \subset  \! \{0,1\}^{\dimension}$ & &
    $\setB \! = \! \{\elementB_1, \ldots, \elementB_{\sizeB}\} \! \subset \! \{0,1\}^{\dimension}$ \\
    \cline{2-2} \cline{4-4}
    \li & $I \gets \emptyset$ \\
    \li & \kwfor $i \in \range{\sizeA}$: & & \kwfor $i \in \range{\sizeA}$: \\
    \li[shade-start] & \ind $\vv{D_i} \gets 0^{\sizeB}$ & & \ind $\vv{M_i} \gets 0^{\sizeB}$ \\
    \li & \ind \kwfor $j \in \range{\dimension}$: & &
    \ind \kwfor $j \in \range{\dimension}$: \\
    \li & \indd $c \gets q_i[j]$ & &
    \indd $\vv{m} \sample \group^{\sizeB}$ \\
    \li & & & $\indd \vv{r} \gets (r_1[j], \cdots, r_{\sizeB}[j])^T$ \\
    \li & & & \indd $\vv{m_0} \getsp \vv{m} + r $ \\
    \li & & & \indd $\vv{m_1} \getsp \vv{m} + (1^{\sizeB} \xor \vv{r}) $ \\
    \li & \indd $\vv{m_{i,j}} \gets \textsf{OTrecv}(c)$ & &
    \indd $\textsf{OTsend}(\vv{m_0}, \vv{m_1})$ \\
    \li[shade-end] & \indd $\vv{D_i} \getsp \vv{D_i} + \vv{m_{i,j}}$ & &
    \indd $\vv{M_i} \getsp \vv{M_i} + \vv{m}$ \\
    \li & \ind \kwfor $k \in \range{\sizeB}$: & &
    \ind \kwfor $k \in \range{\sizeB}$: \\
    \li & & & \indd \kwfor $m \in \group$: \\
    \li & & & \inddd $v_m \! \gets \! (m \! - \! \vv{M_i}[k] \bmod{p}) \! \leq \! \tau$ \\
    \li & \indd $b_{i,k} \gets \textsf{OTrecv}(\vv{D_i}[k])$ & &
    \indd $\textsf{OTsend}(v_0, \ldots, v_{p - 1})$ \\
    \li & \indd \kwif $b_{i,k} = 1$:\\
    \li & \inddd $I \gets I \cup \{ (i,k) \}$ \\
    \li & \kwreturn $I$
    \end{tabular}
    \caption{Full \otfpsi protocol. Lines \lineref{otfpsi}{shade-start}-\lineref{otfpsi}{shade-end} are the SHADE construction.}
    \label{fig:fpsi:otfpsi}
\end{figure}

\descr{Batching.} When computing the distances between one element \elementA held by \partyA and all elements in \setB, \partyA's input does not change (i.e., it is always $\elementA[i]$ for the $i$-th bit computation). 
Following SHADE~\cite{Bringer2013}, we batch these into one (correlated) OT. This strategy reduces the number of OTs from $\sizeB\dimension$ to $\dimension$. While it requires larger OT messages, this can be achieved inexpensively using pseudo-random functions (see \cref{ap:ot}). This batching allows us to reduce computation cost.

\descr{Full Construction.} We present the full \otfpsi protocol in \cref{fig:fpsi:otfpsi}. 
To compare two sets \setA and \setB, it loops over each $\elementA \in \setA$, computes the distances to all $\elementB \in \setB$ using batching, and compares each computed distance to the threshold \threshold. We proof correctness and security of \otfpsi in \cref{ap:proofs}.

\descr{Complexity.} We analyze the asymptotic complexity of \otfpsi.
The distance computation step performs $\sizeA\dimension$ 1-out-of-2 chosen OTs with a message length of $\sizeB\log \modulus$. As $\modulus \in \bigO{l}$, this results in a communication and computation complexity of \bigO{\sizeA\sizeB\dimension\log\dimension}. The threshold comparison step performs $\sizeA\sizeB$ 1-out-of-\modulus chosen 1-bit OTs, resulting in a communication and computation complexity of \bigO{\sizeA\sizeB\dimension}.
Assuming $\sizeA, \sizeB \in \bigO{n}$, \otfpsi is quadratic in $n$. This is asymptotically worse than existing protocols that have communication (and, for Fmap-FPSI \cite{Gao2024}, computation) only (quasi-)linear in $n$. However, these protocols only achieve this by relying on restrictive input assumptions \cite{Chongchitmate2024,Gao2024} or suboptimal complexities in the dimension \cite{Gao2024} or threshold \cite{Gao2024,Blass2025}. In \S\ref{sec:eval:rw}, we show that \otfpsi is concretely more efficient than these protocols for most practical parameters.

\subsection{Secret-Shared \otfpsi}
\label{subsec:fpsi:secret_shared}

As described in \S\ref{sec:system}, \name relies on a secret-shared FPSI protocol (\cref{fig:sytem:ssfpsi_ideal}). This allows \partyA and \partyB to outsource the computation and communication cost of the FPSI protocol to two non-colluding compute nodes \serverA and \serverB. To do so, \partyA and \partyB generate bitwise secret-shares of their input sets $\setA = \shareA{\setA} \xor \shareB{\setA}$ and $\setB = \shareA{\setB} \xor \shareB{\setB}$. Both parties then send one share to each of the two non-colluding nodes, which then run a secret-shared FPSI protocol. \partyA can retrieve the secret-shares of the result from the nodes and reconstruct the result.

In this section, we describe \otfpsiss and \otfpsissb, two secret-shared variants of \otfpsi. 

\descr{Single Comparison.} For a single comparison, operating on bitwise secret shares is straightforward: Assume \serverA holds secret shares \shareA{\elementA}, \shareA{\elementB} and \serverB holds \shareB{\elementA}, \shareB{\elementA} where $ \shareA{\elementA} \xor \shareB{\elementA} = \elementA$ and $\shareA{\elementB} \xor \shareB{\elementB} = \elementB $. Observe that $\hammingdist(\elementA, \elementB) = \hammingweight(\elementA \xor \elementB)= \hammingweight(\shareA{\elementA} \xor \shareA{\elementB} \xor \shareB{\elementA} \xor \shareB{\elementB}) = \hammingdist(\shareA{\elementA} \xor \shareA{\elementB}, \shareB{\elementA} \xor \shareB{\elementB})$ where \hammingweight denotes the Hamming weight. Thus, \serverA and \serverB can locally XOR their shares $\shareA{\elementA} \xor \shareA{\elementB}$ and $\shareB{\elementA} \xor \shareB{\elementB}$, and invoke the private comparison protocol from \otfpsi. 
To create secret-shared outputs, we modify the threshold comparison as follows: \serverB samples a random bit $\shareB{P}$ and uses it to mask the comparison results: $v_i = (i - M \mod \modulus \le \threshold) \xor \shareB{P}$ for $i \in \group$. Server \serverA retrieves $\shareA{P} = v_D$ through OT and outputs $\shareA{P}$, \serverB outputs $\shareB{P}$. 

\subdescr{Correctness.} %
By construction, we have $D - M \mod p = \hammingdist(\shareA{\elementA} \xor \shareA{\elementB}, \shareB{\elementA} \xor \shareB{\elementB}) = \hammingdist(\elementA, \elementB)$ and $\shareA{P} = (D - M \mod \modulus \le \threshold) \xor \shareB{P} = (\hammingdist(\elementA, \elementB) \leq \threshold) \xor \shareB{P}$, hence $\shareA{P} \xor \shareB{P} = (\hammingdist(\elementA, \elementB) \le \threshold)$.

\descr{\otfpsiss.} Our first secret-shared FPSI protocol, \otfpsiss, applies the single comparison outlined above to all pairs of records across \setA and \setB. \cref{fig:fpsi:otfpsiss} presents the full protocol. We prove correctness and security in \cref{ap:proofs}.

\begin{figure}
    \footnotesize
    \setlisting{otfpsiss}
    \setcounter{linecounter}{1}
    \renewcommand{\arraystretch}{1.3}
    \begin{tabular}{@{}r@{}lc@{}l@{}}
        & \textbf{Server \serverA} & & \textbf{Server \serverB} \\
        & $\shareA{\setA} = \{\shareA{\elementA_1}, \ldots, \shareA{\elementA_{\sizeA}}\} \! \subset \! \{0,1\}^{\dimension}$ & & $\shareB{\setA} = \{\shareB{\elementA_1}, \ldots, \shareB{\elementA_{\sizeA}}\} \! \subset \! \{0,1\}^{\dimension}$ \\
        & $\shareA{\setB} = \{\shareA{\elementB_1}, \ldots, \shareA{\elementB_{\sizeA}}\} \! \subset \! \{0,1\}^{\dimension}$ & & $\shareB{\setB} = \{\shareB{\elementB_1}, \ldots, \shareB{\elementB_{\sizeA}}\} \! \subset \! \{0,1\}^{\dimension}$ \\
        \cline{2-2} \cline{4-4}
        \li & $D, \shareA{P} \gets 0^{\sizeA\times\sizeB}$ & & $M, \shareB{P} \gets 0^{\sizeA\times\sizeB}$ \\ 
        \li & \kwfor $i \in \range{\sizeA}, j \in \range{\sizeB}$: & & \kwfor $i \in \range{\sizeA}, j \in \range{\sizeB}$ \\
        \li & \ind \kwfor $k \in \range{\dimension}$: & & \ind \kwfor $k \in \range{\dimension}$: \\      
        \li & \indd $ \shareA{b} \gets \shareA{\elementA_i}[k] \xor \shareA{\elementB_j}[k] $ & & \indd $\shareB{b} \gets \shareB{\elementA_i}[k] \xor \shareB{\elementB_j}[k] $ \\
        \li & & & \indd $m \sample \group$ \\
        \li & & & \indd $m_0 \getsp m + \shareB{b}$ \\
        \li & & & \indd $m_1 \getsp m + (1\xor \shareB{b})$ \\
        \li & \indd $m_{i,j,k} \gets \textsf{OTRecv}(\shareA{b})$ & & \indd $\textsf{OTSend}(m_0, m_1)$ \\
        \li & \indd $D_{i,j} \getsp D_{i,j} + m_{i,j,k}$ & & \indd $M_{i,j} \getsp M_{i,j} + m$ \\
        \li & & & \ind $\shareB{P}_{i,j} \sample \bin$ \\
        \li & & & \ind \kwfor $m \in \group$: \\
        \li & & & \indd $v_m \gets (m-M_{i,j} \bmod \modulus \leq \threshold)$ \\
        \li & & & \indd $v_m \gets v_m \xor \shareB{P}_{i,j}$ \\
        \li & \ind $\shareA{P}_{i,j} \gets \textsf{OTrecv}(M_{i,j})$  & & \ind $\textsf{OTSend}(v_0, \dots, v_{\modulus-1})$ \\
        \li & \kwreturn $\shareA{P}$ & & \kwreturn $\shareB{P}$ \\ 
    \end{tabular}
    \caption{Full \otfpsiss protocol.}
    \label{fig:fpsi:otfpsiss}
\end{figure}

\subdescr{Batching. } We cannot apply the same batching strategy as in \otfpsi to the secret-shared setting. In \otfpsi, \partyA's OT inputs are determined by \elementA only and are the same when comparing one \elementA to any $\elementB \in \setB$. In \otfpsiss, \serverA's OT inputs are determined by $\shareA{\elementA} \xor \shareA{\elementB}$, which differs for different $\elementB \in \setB$. 

\descr{\otfpsissb.} Performing many OTs is expensive (although the choice of OT may allow a trade-off between communication and computation). Our second secret-shared FPSI protocol, \otfpsissb, utilizes a different batching approach to reduce the number of OTs from $\sizeA\sizeB\dimension$ to $(\sizeA+\sizeB)\dimension$ at the cost of additional communication. \otfpsissb can concretely reduce cost for $\sizeA > 1$ (see \S\ref{subsec:eval:secret_shared}). 

\subdescr{Using 1-out-of-4 OT.} The \otfpsissb protocol relies on 1-out-of-4 OT for the distance computation step: When comparing the $k$-th bit of the secret-shared bit strings \elementA and \elementB, both parties run a 1-out-of-4 OT into which \serverA inputs the two bits $\shareA{\elementA}[k]$ and $\shareA{\elementB}[k]$ individually instead of their XOR. 

As before, \serverB samples a random mask $m \sample \group$ and computes four OT messages $m_0,\dots,m_3$. \serverA chooses the message indexed by $c = 2\shareA{\elementA}[k] + \shareA{\elementB}[k]$. As for plaintext comparison (\S\ref{subsec:fpsi:single_comparison}), we want that $m_c - m \bmod \modulus = \elementA[k] \xor \elementB[k] = \shareA{\elementA}[k] \xor \shareA{\elementB}[k] \xor \shareB{\elementA}[k] \xor \shareB{\elementB}[k]$. We can achieve this by setting the four OT messages for $\shareB{b} = \shareB{\elementA}[k] \xor \shareB{\elementB}[k]$ as $m_0 = m_3 = m + \shareB{b} \mod p$ and $m_1=m_2 = m + (1 \xor \shareB{b}) \mod p$. 

\subdescr{Correctness.} If $\shareA{\elementA}[k] \xor \shareA{\elementB}[k] = 0$, we have $m_c-m \mod \modulus = \shareB{b} =  \shareB{\elementA}[k] \xor \shareB{\elementB}[k] = \shareA{\elementA}[k] \xor \shareA{\elementB}[k] \xor \shareB{\elementA}[k] \xor \shareB{\elementB}[k] = \elementA[k] \xor \elementB[k]$.  If $\shareA{\elementA}[k] \xor \shareA{\elementB}[k] = 1$, then $m_c-m \mod \modulus = 1 \xor \shareB{b} = 1 \xor \shareB{\elementA}[k] \xor \shareB{\elementB}[k] = \shareA{\elementA}[k] \xor \shareA{\elementB}[k] \xor \shareB{\elementA}[k] \xor \shareB{\elementB}[k] = \elementA[k] \xor \elementB[k]$.

\subdescr{Naor-Pinkas construction.} The Naor-Pinkas construction \cite{Naor1999} allows us to implement a \emph{random} 1-out-of-4 OT running two independent random 1-out-of-2 OTs. In random OT, \serverB does not choose the OT messages, but learns the randomly chosen messages $\omega_0,\dots,\omega_3$ during the protocol. 
More precisely, for \serverA's choice $c = 2\shareA{\elementA}[k] + \shareA{\elementB}[k]$, both parties run one random 1-out-of-2 OT for each input bit $\shareA{\elementA}[k]$ and $\shareA{\elementB}[k]$. In the first OT, \serverB learns two random messages $\alpha_0,\alpha_1$, and \serverA learns $\alpha_{\shareA{\elementA}[k]}$. In the second OT, \serverB learns $\beta_0,\beta_1$ and \serverA learns $\beta_{\shareA{\elementB}[k]}$. 
Using a family of pseudo-random functions $F_k: \bin^* \rightarrow \group$ for $k \in \bin^\secpar$, \serverB can compute the four random OT messages $\omega_0, \dots, \omega_3$ as $\omega_j = F_{\alpha_{j_1}}(j) + F_{\beta_{j_2}}(j) \mod \modulus$ where $j = 2j_1+j_2$. By construction, \partyA can only compute $\omega_c = F_{\alpha_{\shareA{\elementA}[k]}}(j) + F_{\beta_{\shareA{\elementB}[k]}}(c)$. 

\subdescr{Correlated OT.}  As with the plaintext comparison, correlated OT (see \cref{ap:ot}) can be used instead of chosen OT. Let $m_0 \in\group$ be the random message chosen by correlated OT. Then, we set $m = m_0 - \shareB{b} \mod \modulus$ and compute the remaining messages as 
$m_3 = m_0$  and $m_1=m_2 = m + (1 \xor \shareB{b}) \bmod \modulus$.

The Naor-Pinkas construction provides a random 1-out-of-4 OT. To implement chosen OT, \serverB can use the random messages to mask its actual messages and send them to \serverA. Implementing correlated OT is cheaper and can be done by only sending three messages: Let $\omega_1, \dots, \omega_3 \in \group$ be the random OT messages. \serverB sets $m_0 = \omega_0$ and computes $m_1,m_2,m_3$ as above. Then, \serverB masks the $m_1,m_2,m_3$ as $\mu_i = m_i - \omega_i \mod \modulus$ and sends $\mu_1,\mu_2,\mu_3$ to \serverA, which can unmask $m_c = \mu_c + \omega_c \mod \modulus$ where $\mu_0 = 0$.

\subdescr{The Key Observation.} With the Naor-Pinkas construction, we can compare two secret-shared bit strings by running individual and independent OTs for each secret share held by \serverA. When dealing with two secret-shared sets \setA and \setB instead of two strings, we observe that for all comparisons of a specific $\elementA \in \setA$ to any $\elementB \in \setB$, \serverA's input in the first OT of the Naor-Pinkas construction for the $k$-th bit is always $\shareA{\elementA}[k]$. Similarly, when comparing a specific $\elementB \in \setB$ to any $\elementA \in \setA$, \partyA's input to the second OT for the $k$-th bit is always $\shareB{\elementB}[k]$.

This key observation allows us to reduce the number of random 1-out-of-2 OTs we need: Instead of running one OT for each bit of every comparison (as \otfpsiss), we only need one OT for each bit of every input share -- which is an improvement for $\sizeA,\sizeB > 1$.

\begin{figure}
    \footnotesize
    \setlisting{otfpsissb}
    \setcounter{linecounter}{1}
    \renewcommand{\ind}{\;\;\;}
    \renewcommand{\indd}{\;\;\;\;\;\;}
    \renewcommand{\arraystretch}{1.3}
    \begin{tabular}{@{}r@{}lc@{}l@{}}
        & \textbf{Server \serverA} & & \textbf{Server \serverB} \\
        & $\shareA{\setA} = \{\shareA{\elementA_1}, \ldots, \shareA{\elementA_{\sizeA}}\} \! \subset \! \{0,1\}^{\dimension}$ & & $\shareB{\setA} = \{\shareB{\elementA_1}, \ldots, \shareB{\elementA_{\sizeA}}\} \! \subset \! \{0,1\}^{\dimension}$ \\
        & $\shareA{\setB} = \{\shareA{\elementB_1}, \ldots, \shareA{\elementB_{\sizeA}}\} \! \subset \! \{0,1\}^{\dimension}$ & & $\shareB{\setB} = \{\shareB{\elementB_1}, \ldots, \shareB{\elementB_{\sizeA}}\} \subset \{0,1\}^{\dimension}$ \\
        \cline{2-2} \cline{4-4}
        \li & $D, \shareA{P} \gets 0^{\sizeA \times \sizeB}$ & & $M, \shareB{P} \gets 0^{\sizeA \times \sizeB}$  \\
        \li & \kwfor $k \in \range{\dimension}$: & & \kwfor $k \in \range{\dimension}$:  \\
        \li[start_ots] & \ind \kwfor $i \in \range{\sizeA}$: & & \ind \kwfor $i \in \range{\sizeA}$: \\
        \li & & & \indd $X_{i,k}^0, X_{i,k}^1 \gets \bin^\secpar$ \\
        \li & \indd $X_{i,k} \gets \textsf{OTRecv}(\shareA{\elementA_i}[k])$ & & \indd $\textsf{OTSend}(X_{i,k}^0,X_{i,k}^1)$ \\
        \li & \ind \kwfor $j \in \range{\sizeB}$: & & \ind \kwfor $j \in \range{\sizeB}$: \\
        \li & & & \indd $Y_{j,k}^0, Y_{j,k}^1 \gets \bin^\secpar$ \\
        \li[end_ots] & \indd $Y_{j,k} \gets \textsf{OTRecv}(\shareA{\elementB_j}[k])$ & & \indd $\textsf{OTSend}(Y_{j,k}^0,Y_{j,k}^1)$ \\
        \li & \ind \kwfor $i \in \range{\sizeA}, j \in \range{\sizeB}$: & & \ind \kwfor $i \in \range{\sizeA}, j \in \range{\sizeB}$: \\
        \li & \indd $z \gets (i,j,k)$ & & \indd $z \gets (i,j,k)$ \\
        \li & \indd $c_z \gets 2\shareA{\elementA_i}[k] + \shareA{\elementB_j}[k]$& & \indd $\shareB{b}_z \gets \shareB{\elementA_i}[k] \xor \shareB{\elementB_j}[k]$ \\
        \li[np_start] & & & \indd \kwfor $x = (x_1,x_0) \in \{0, \dots, 3\}$:\\
        \li[f_z] & \indd $f_z \getsp \prfff{c_z}{X_{i,k}}$ & & \inddd $\omega_{z,x} \getsp \prfff{x}{X_{i,k}^{x_1}}$ \\
        \li[np_end] & \indd $f_z \getsp f_z + \prfff{c_z}{Y_{j,k}}$ & & \inddd $\omega_{z,x} \getsp \omega_{z,x} + \prfff{x}{Y_{j,k}^{x_0}}$ \\
        \li[m_start] & & & \indd $m_{z,0}, m_{z,3} \getsp \omega_{z,0}$ \\
        \li & & & \indd $m_z \getsp \omega_{z,0} - \shareB{b}_z$ \\
        \li[m_end] & & & \indd $m_{z,1}, m_{z,2} \getsp m_{z} + (1 \xor \shareB{b}_z)$ \\ 
        \li[mask] & & & \indd \kwfor $x \in \{1, \dots, 3\}$:\\
        \li & & & \inddd $\mu_{z,x} \getsp m_{z,x} - \omega_{z_x} $ \\
        \li[send_mus] & \indd $\mu_{z,1}, \mu_{z,2}, \mu_{z,3} \! \gets \! \textsf{Recv}()$ & & \indd $\textsf{Send}(\mu_{z,1},\mu_{z,2},\mu_{z,3})$ \\
        \li[d] & \indd $d_z \getsp f_z + \mu_{z,c_z}$  \\
        \li & \indd $D_{i,j}\getsp D_{i,j} + d_z$ & & \indd $M_{i,j} \getsp M_{i,j} + m_z$ \\
        \li & \kwfor $i \in \range{\sizeA}, j \in \range{\sizeB}$: & & \kwfor $i \in \range{\sizeA}, j \in \range{\sizeB}$: \\
        \li & & & \ind $\shareB{P}_{i,j} \sample \bin$ \\
        \li & & & \ind \kwfor $m \in \group$: \\
        \li & & & \indd $v_m \gets (m - M_{i,j} \bmod \modulus \leq \threshold)$ \\
        \li & & & \indd $v_m \gets v_m \xor \shareB{P}_{i,j}$ \\
        \li & \ind $\shareA{P}_{i,j} \gets \textsf{OTRecv}(M_{i,j})$ & & \ind $\textsf{OTSend}(v_0, \dots, v_{\modulus - 1})$ \\
        \li & \kwreturn $\shareA{P}$ & & \kwreturn $\shareB{P}$
        \end{tabular}
    \caption{Full \otfpsissb protocol.}
    \label{fig:fpsi:otfpsissb}
\end{figure}

\subdescr{The Full Protocol.} \Cref{fig:fpsi:otfpsissb} presents the full protocol. For every bit, we first run one (random) OT for every share held by \serverA (lines~\lineref{otfpsissb}{start_ots}-\lineref{otfpsissb}{end_ots}), resulting in the random seeds $X_i$ for $i \in \range{\sizeA}$ and $Y_j$ for $j \in \range{\sizeB}$. For each bit comparison $z$, \serverB derives the random OT messages $\omega_{z,x}$ (lines~\lineref{otfpsissb}{np_start}-\lineref{otfpsissb}{np_end}) and computes the $m_{z,x}$ as outlined above (lines~\lineref{otfpsissb}{m_start}-\lineref{otfpsissb}{m_end}).

Afterwards, \serverB masks the other three OT messages using the random Naor-Pinkas OT messages and sends these masked values to \serverA (lines~\lineref{otfpsissb}{mask}-\lineref{otfpsissb}{send_mus}), who reconstructs $d = m_c$ (lines~\lineref{otfpsissb}{f_z}, \lineref{otfpsissb}{np_end}, \lineref{otfpsissb}{d}). After computing the distances for all pairs of bit strings, \serverA and \serverB run the same secret-shared threshold comparison protocol as in \otfpsiss.
We prove the correctness and security of \otfpsissb in \cref{ap:proofs}.

\section{Evaluation}
\label{sec:evaluation}
\begin{table*}
    \centering
    \caption{Run time and communication of \otfpsi using SilentOT for querier set size \sizeA, responder set size \sizeB, and dimension $\dimension$ (threshold $\tau = \floor{\dimension/16}$). %
    }
    \label{tab:eval:otfpsi_performance_silent}
    \begin{tabular}{rrrrrrrrrrr}
        \toprule
        & & \multicolumn{3}{c}{$\dimension = 127$} & \multicolumn{3}{c}{$\dimension = 511$} & \multicolumn{3}{c}{$\dimension = 8191$}  \\
        \cmidrule(lr){3-5} \cmidrule(lr){6-8} \cmidrule(lr){9-11}
        \sizeA & \sizeB & Gigabit & Slow & Comm & Gigabit & Slow & Comm & Gigabit & Slow & Comm\\
        \midrule
        \num{64} & \num{64} & \qty{0.031}{\second} & \qty{0.862}{\second} & \qty{0.622}{\mebi\byte} & \qty{0.080}{\second} & \qty{1.005}{\second} & \qty{2.668}{\mebi\byte} & \qty{1.240}{\second} & \qty{4.143}{\second} & \qty{68.209}{\mebi\byte} \\
        \num{256} & \num{256} & \qty{0.278}{\second} & \qty{1.341}{\second} & \qty{8.260}{\mebi\byte} & \qty{1.046}{\second} & \qty{2.672}{\second} & \qty{40.745}{\mebi\byte} & \qty{24.716}{\second} & \qty{49.067}{\second} & \qty{1.063}{\gibi\byte} \\
        \num[round-mode=none,exponent-mode=input]{1024} & \num[round-mode=none,exponent-mode=input]{1024} & \qty{4.409}{\second} & \qty{7.393}{\second} & \qty{129.965}{\mebi\byte} & \qty{16.474}{\second} & \qty{29.570}{\second} & \qty{649.386}{\mebi\byte} & \qty{418.093}{\second} & \qty{895.485}{\second} & \qty{17.001}{\gibi\byte} \\
        \num[round-mode=none,exponent-mode=input]{4096} & \num[round-mode=none,exponent-mode=input]{4096} & \qty{70.101}{\second} & \qty{103.130}{\second} & \qty{2.028}{\gibi\byte} & \qty{266.900}{\second} & \qty{461.935}{\second} & \qty{10.143}{\gibi\byte} & \qty[round-precision=4]{6805.789}{\second} & \qty{15036.354}{\second} & \qty{271.998}{\gibi\byte} \\
        \midrule
        \num{1} & \num[round-mode=none,exponent-mode=input]{16384} & \qty{0.078}{\second} & \qty{1.024}{\second} & \qty{2.128}{\mebi\byte} & \qty{0.261}{\second} & \qty{1.417}{\second} & \qty{10.257}{\mebi\byte} & \qty{4.602}{\second} & \qty{13.684}{\second} & \qty{272.186}{\mebi\byte} \\
        \num{1} & \num[round-mode=none,exponent-mode=input]{131072} & \qty{0.536}{\second} & \qty{1.700}{\second} & \qty{16.336}{\mebi\byte} & \qty{1.977}{\second} & \qty{4.705}{\second} & \qty{81.270}{\mebi\byte} & \qty{36.534}{\second} & \qty{103.180}{\second} & \qty{2.125}{\gibi\byte} \\
        \num{1} & \num[round-mode=none,exponent-mode=input]{524288} & \qty{2.118}{\second} & \qty{4.116}{\second} & \qty{65.009}{\mebi\byte} & \qty{7.812}{\second} & \qty{15.228}{\second} & \qty{324.704}{\mebi\byte} & \qty{146.291}{\second} & \qty{366.747}{\second} & \qty{8.500}{\gibi\byte} \\
        \num{1} & \num[round-mode=none,exponent-mode=input]{1048576} & \qty{4.210}{\second} & \qty{7.426}{\second} & \qty{129.897}{\mebi\byte} & \qty{15.604}{\second} & \qty{29.767}{\second} & \qty{649.272}{\mebi\byte} & \qty{292.859}{\second} & \qty{726.060}{\second} & \qty{17.000}{\gibi\byte} \\
        \bottomrule
    \end{tabular}
\end{table*}

\descr{Implementation. }
To demonstrate the performance of \name, we implement the core FPSI construction in C++ and provide extensive benchmarks. We publish this implementation as part of our artifact \cite{rausch2025xdupARTIFACT}.
For 1-out-of-2 OT, we use SilentOT \cite{Boyle2019} provided by the libOTe library \cite{Rindal2025} (in \cref{ap:eval:soft}, we also provide evaluations with SoftSpokenOT \cite{Roy2022}). 
We implement the 1-out-of-\dimension OT required for the distance comparison from 1-out-of-2 OT using the Naor-Pinkas construction~\cite{Naor1999}. This approach proved more efficient than 1-out-of-N OT \cite{Kolesnikov2016} as \dimension is relatively small, and we only need 1-bit messages. We implement the PRF using AES-CTR. 

\descr{Environment. }
Prior \ac{FPSI} works often benchmark their protocols in high-resource environments \cite{Uzun2021,Chakraborti2023,Gao2024}. To showcase the practical performance of \otfpsi, we deliberately choose a relatively low-resource environment: All our experiments run on a single Google Cloud Platform C4D VM with 4 cores of an AMD EPYC Turin CPU and \qty{30}{\giga\byte} of RAM. Both parties run the computation single-threaded.

While our approach is computation-efficient, it has a comparatively high communication cost. To provide a meaningful evaluation, we simulate two network connections: a high-quality LAN with \qty{1}{\giga\bit\per\second} and \qty{.5}{\milli\second} latency, and a slower connection of \qty[exponent-mode=input]{250}{\mega\bit\per\second} with \qty{20}{\ms} latency. 
For a fair comparison with prior work, we match their respective network conditions.
As we observed little variance in preliminary runs, we report numbers from single runs. 

\subsection{Performance of \otfpsi}\label{sec:eval:otfpsi}

\cref{tab:eval:otfpsi_performance_silent} shows the runtime and communication cost of \otfpsi with SilentOT for a symmetric setting where both parties hold a set of the same size, and an asymmetric setting where the querier only holds one record. We add a large dimension for comparison, yet our \name does not need $\dimension=8191$. Our results confirm that run time and communication of \otfpsi are linear in the number of comparisons $\sizeA\sizeB$. We confirm that the network setting influences the run time of \otfpsi, as it is relatively communication-heavy.

\subsection{Comparison to Existing FPSI Protocols} 
\label{sec:eval:rw}

\begin{table}    
    \centering
    \caption{Online and total computation time of FLPSI~\cite{Uzun2021} (excluding communication, sub-sampling parameters $t=2$, $T =64$) and total run time of \otfpsi ($\sizeA = 1$, $\threshold=25$) over gigabit and slow network with SilentOT ($\dimension=256$).}
    \label{tab:comp_fpsi_silentot}
    \begin{tabular}{lrrrrrr}
        \toprule
         & \multicolumn{3}{c}{FLPSI} & \multicolumn{3}{c}{\otfpsi} \\
        \cmidrule(lr){2-4} \cmidrule(lr){5-7}
        \sizeB & Online & Total & Comm & Gigabit & Slow & Comm \\
        \midrule
        $10^4$ & \qty{0,523}{\second} & \qty{1,463}{\second} & \qty{12.1}{\mebi\byte} & {\bfseries \qty[text-series-to-math]{0.112}{\second}} & \qty{1.063}{\second} & \qty{3.216}{\mebi\byte}\\
        $10^5$ & \qty{4,4457}{\second} & \qty{8,527}{\second} & \qty{20.4}{\mebi\byte} & {\bfseries \qty[text-series-to-math]{0.825}{\second}} & \qty{2.455}{\second} & \qty{31.200}{\mebi\byte} \\
        $10^6$ & \qty{43,956}{\second} & \qty{81,456}{\second} & \qty{40.8}{\mebi\byte} & {\bfseries \qty[text-series-to-math]{8.720}{\second}} & \qty{16.046}{\second} & \qty{310.913}{\mebi\byte} \\
        \bottomrule
    \end{tabular}
\end{table}

\begin{figure}
    \centering
    \subfloat{
        \label{fig:eval:comp_da_approx_dimension}
        \includegraphics[width=0.45\columnwidth]{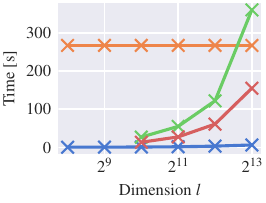}
    }
    \hfill
    \subfloat{
        \label{fig:eval:comp_da_approx_threshold}
        \includegraphics[width=0.45\columnwidth]{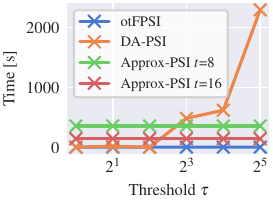}
    }
    \caption{Runtime \otfpsi with SilentOT, DA-PSI, and Approx-PSI ($\sizeA = \sizeB = 100$, \qty{320}{\mega\bit\per\second}, \qty{20}{\milli\second} latency).}
    \label{fig:eval:comp_da_approx}
\end{figure}

\begin{table}
    \centering
    \caption{Run time and communication of Approx-PSI~\cite{Chongchitmate2024} (gap $t = \log \dimension$) and \otfpsi with SilentOT by set size $n = \sizeA = \sizeB$ ($\dimension=128$, $\threshold=4$, gigabit network).}
    \label{tab:eval_comp_approx_set_size_silent}
    \begin{tabular}{rrrrr}
        \toprule
        & \multicolumn{2}{c}{Approx-PSI} & \multicolumn{2}{c}{\otfpsi} \\
        \cmidrule(lr){2-3} \cmidrule(lr){4-5}
        $n$ & Run time & Comm & Run time & Comm \\
        \midrule
        \num[exponent-mode=input]{256} & \qty{38.7}{\second} & \qty{465.68}{\mebi\byte} & {\bfseries \qty[text-series-to-math]{0.310}{\second}} & \qty{9.219}{\mebi\byte} \\
        \num[exponent-mode=input,round-mode=none]{1024} & \qty{147.85}{\second} & \qty{1,737597656}{\gibi\byte} & {\bfseries \qty[text-series-to-math]{4.595}{\second}} & \qty{145.324}{\mebi\byte}  \\
        \num[exponent-mode=input,round-mode=none]{4096} & \qty{569,9}{\second} & \qty{6,708984375}{\gibi\byte} & {\bfseries \qty[text-series-to-math]{72.805}{\second}} & \qty{2.268}{\gibi\byte} \\
        \bottomrule
    \end{tabular}
\end{table}

\begin{table}[tbp]
    \centering
    \caption{Run time of Fmap-FPSI~\cite{Gao2024} and \otfpsi with SilentOT by (a) threshold \threshold ($\dimension=512$, $\sizeA=\sizeB=512$) and (b) dimension \dimension ($\threshold=\dimension/16$, $\sizeA=\sizeB=128$). \qty{10}{\gibi\bit\per\second}, \qty{.02}{\milli\second} latency.}
    \label{tab:eval:comp_fmap_fpsi_threshold_dimension_silent}
    \begin{tabular}{lrrrrrr}
        \toprule
        && \multicolumn{3}{c}{Fmap-FPSI} & \multicolumn{2}{c}{\otfpsi} \\
        \cmidrule(lr){3-5} \cmidrule(lr){6-7}
        & \threshold{} / \dimension & Online & Total & Comm & Total & Comm \\
        \midrule
        \threshold & 1 & {\bfseries \qty{1.186}{\second}} & \qty{316.249}{\second} & \qty{292.916}{\mebi\byte} & \qty{4.583}{\second} & {\qty{187.220}{\mebi\byte}} \\
        & 2 & {\bfseries \qty{1.909}{\second}} & \qty{475.831}{\second} & \qty{439.479}{\mebi\byte} & \qty{4.577}{\second} & {\qty{187.220}{\mebi\byte}} \\
        & 4 & {\bfseries \qty{3.238}{\second}} & \qty{794.930}{\second} & \qty{732.769}{\mebi\byte} & \qty{4.588}{\second} & {\qty{187.220}{\mebi\byte}} \\
        & 8 & \qty{28.032}{\second} & \qty{1458.342}{\second} & \qty{1,288330078}{\gibi\byte} & {\bfseries \qty{4.578}{\second}} & {\qty{187.220}{\mebi\byte}} \\
        & 16 & \qty{64.813}{\second} & \qty{2971.311}{\second} & \qty{2,73215332}{\gibi\byte} & {\bfseries \qty{4.586}{\second}} & {\qty{187.220}{\mebi\byte}} \\
        & $\geq$ 32 & \multicolumn{3}{c}{Unsupported parameters.} & {\bfseries \qty{4.575}{\second}} & {\qty{187.220}{\mebi\byte}} \\

        \midrule
        \dimension & 64 & \qty{0.941593}{\second} & \qty{30.439}{\second} & \qty{30.305}{\mebi\byte} & {\bfseries \qty{0.067}{\second}} & \qty{1.155}{\mebi\byte} \\
        & 128 & \qty{3.481}{\second} & \qty{115.064}{\second} & \qty{116.363}{\mebi\byte} & {\bfseries \qty{0.093}{\second}} & \qty{2.398}{\mebi\byte}  \\
        & 256 & \qty{12.917}{\second} & \qty{447.227}{\second} & \qty{455.452}{\mebi\byte} & {\bfseries \qty{0.175}{\second}} & \qty{5.234}{\mebi\byte} \\
        & 512 & \qty{48.632}{\second} & \qty{1761.740}{\second} & \qty{1801.254}{\mebi\byte} &{\bfseries \qty{0.308}{\second}} & \qty{11.840}{\mebi\byte}  \\
        & 1024 & \multicolumn{3}{c}{Ran out of memory.} & {\bfseries \qty{0.588}{\second}} & \qty{27.788}{\mebi\byte} \\
        \bottomrule
    \end{tabular}
\end{table}

To validate the performance of \otfpsi, we compare it to prior FPSI protocols. 
Except for Fmap-FPSI \cite{Gao2024}, the code of existing FPSI protocols was not public at the time of writing. Hence, we can only compare to the numbers the authors report in the respective paper. For a fair comparison, we match protocol parameters and network setting. 
While we cannot replicate the original hardware, we use a relatively low-resource environment compared to the environments of prior work (Appendix, \cref{tab:eval_rw_environments}).

We evaluate and compare at different dimensions and thresholds. Yet, for our humanitarian use case, we aim for a dimension $\dimension \approx 512$ and a threshold of $\threshold \approx \dimension/4$ (\cref{sec:embedding}).

\descr{FLPSI~\cite{Uzun2021}.} We compare the run time of \otfpsi (including communication) to the computation time of FLPSI (excluding communication) and present the results in \cref{tab:comp_fpsi_silentot}. We observe that \otfpsi has consistently better run times, even when run over slow networks: For a database size of \num[exponent-mode=input]{1000000}, \otfpsi achieves a reduction of 
\qty{9.3}{\x} compared to total computation and 
\qty{5.0}{\x} compared to online computation.

Performance is not the only advantage of \otfpsi: It provides exact results, while FLPSI relies on sub-sampling to approximate the Hamming distance which is fundamentally unable to provide sufficient accuracy (see \cref{ap:comp_FLPSI}).

Bui and Cong~\cite{Bui2025} build an \ac{fpsi} protocol on the same sub-sampling approach, suffering from the same limitations. Still, their protocol is significantly slower than \otfpsi.

\descr{DA-PSI~\cite{Chakraborti2023}.} The Distance-Aware PSI protocol provides a matching protocol with dimension-independent communication cost (but heavily dependent on the threshold \cite[Fig 12]{Chakraborti2023}). To fairly compare to their benchmarks, we evaluate \otfpsi with a \qty[exponent-mode=input]{320}{\mega\bit\per\second} connection and a latency of \qty{10}{\ms}.

\cref{fig:eval:comp_da_approx} shows that \otfpsi is generally faster than DA-PSI. Even for $\dimension = 8192$, \otfpsi is 
\qty{41.7}{\x} faster. 
Additionally, while \otfpsi is not affected by the Hamming distance threshold \threshold, DA-PSI becomes highly impractical for large \threshold: e.g., \otfpsi outperforms DA-PSI by a factor 
\qty{358}{\x}
for $\threshold = 32$. 
Even for small thresholds, their protocol is not efficient enough: We estimate that an offline query at our target set sizes would take 
over \num{17} days even for $\threshold = 4$ -- and DA-FPSI would miss over \qty{90}{\percent} of duplicates when used for deduplication (Appendix, \cref{tab:embedding:smaller_thresh}). %
Finally, DA-FPSI only approximates the distance causing a false-positive rate of \qty{5}{\percent}, violating \reqlinky{fpr}. 

\descr{Approx-PSI~\cite{Chongchitmate2024}.}
The Approx-PSI protocol has a communication and computation complexity near-linear in the set sizes. It assumes that all input data $x, y \in \setA \cup \setB$ either match (i.e., $\hammingweight(x,y) \le \threshold$) or are far apart: $\hammingdist(x,y) \ge t\threshold$ for some gap $t > 3$ (with $t \in \bigO{\log \dimension}$ for near-linear complexity). This assumption can be overly restrictive for large thresholds: For our intended parameters of 
$\threshold \approx \dimension/4$, there is no bit-string set of size three that fulfills this assumption, even for $t = 3$. This assumption is a severe limitation, making Approx-PSI inapplicable to our scenario. For the parameters used by the authors ($\dimension=128$ and $\threshold = 4$), Approx-PSI would miss around \qty{90}{\percent} of duplicates (Appendix, \cref{tab:embedding:smaller_thresh}).
Still, \otfpsi consistently outperforms Approx-PSI (\cref{fig:eval:comp_da_approx}). For gap $t = 8$ and $\dimension=8192$, \otfpsi is faster by a factor of 
\qty{56.4}{\x}. 
\cref{tab:eval_comp_approx_set_size_silent} compares to Approx-PSI for larger sets at their parameterization point: dimension $\dimension = 128$, low threshold $\threshold = 4$, and large gap $t = \log \dimension$. We compare results in our gigabit setting to emulate their LAN.
Even at these parameters, advantageous to Approx-PSI, \otfpsi still outperforms Approx-PSI by
\qty{7.8}{\x}
at a set size of \num[round-mode=none]{4096}. 

\descr{Fmap-FPSI~\cite{Gao2024}.} Fmap-FPSI features both communication and computation linear in the input set sizes by using a new Fuzzy Mapping (Fmap) primitive (which maps elements to a set of IDs such that matching elements will have at least one ID in common). 
Their Fmap relies on a stringent assumption on the input data which limits the threshold \threshold relative to the dimension \dimension. We experimentally evaluate Fmap-FPSI by running the authors' code in our environment and find that Fmap-FPSI does not scale to higher thresholds (\Cref{tab:eval:comp_fmap_fpsi_threshold_dimension_silent}) -- their implementation does not support our target parameters of $\dimension \approx 512$ and $\threshold \approx 128$. \Cref{tab:eval:comp_fmap_fpsi_threshold_dimension_silent} shows that Fmap-FPSI does not scale to large dimensions at $\threshold = \dimension/16$, whereas we are aiming for $\threshold = \dimension/4$. Even for $\dimension=512$ and $\threshold=32$, which Fmap-FPSI only supports for very small sets, the protocol would miss over \qty{73}{\percent} of duplicates when used for deduplication (Appendix, \cref{tab:embedding:smaller_thresh}).  Lastly, Fmap-FPSI relies on an expensive offline phase, making \otfpsi much more competitive even for low thresholds (\Cref{tab:eval:comp_fmap_fpsi_threshold_dimension_silent}).
These observations render Fmap-FPSI less suited for our humanitarian use case (see \cref{ap:comp_FmapFPSI} for more details). 

\descr{PE-FPSI~\cite{Blass2025}.}  The PE-FPSI protocol has linear communication complexity and no input assumptions; it uses predicate encryption. While its communication is asymptotically optimal in the set sizes, its computation is threshold-dependent and concretely inefficient: The largest set sizes evaluated are $\sizeA = \sizeB = 256$. 
Compared to their benchmarks (\num{192} vCPUs, \qty{384}{\gibi\byte} RAM, no latency), and on our more constrained hardware, \otfpsi is still faster than PE-FPSI (see \cref{tab:eval:comp_pe-fpsi}): \otfpsi is
\qty{1529,490616622}{\x}
faster for set size \num{256} and $\threshold = 16$, while reducing communication by 
\qty{43,95789384}{\x}.
Finally, using PE-FPSI for deduplication at the authors' parameters ($\dimension=512$, $\threshold=16$) would miss over \qty{90}{\percent} of duplicates (Appendix, \cref{tab:embedding:smaller_thresh}).

\begin{table}
    \centering
    \caption{Run time and communication of PE-FPSI~\cite{Blass2025} and \otfpsi with SilentOT by set size $n = \sizeA = \sizeB$ ($\dimension=512$, unlimited network)}
    \label{tab:eval:comp_pe-fpsi}
    \begin{tabular}{@{}rrrrrrr@{}}
        \toprule
        & \multicolumn{2}{c}{PE-FPSI ($\threshold = 2$)} & \multicolumn{2}{c}{PE-FPSI ($\threshold = 16$)} & \multicolumn{2}{c}{\otfpsi} \\
        \cmidrule(lr){2-3} \cmidrule(lr){4-5} \cmidrule(lr{0mm}){6-7}
        $n$ & Time & Comm & Time & Comm & Time & Comm \\
        \midrule
        32 & \qty{3.7}{\second} & \qty{35.4}{\mebi\byte} & \qty{28.8}{\second} &  \qty{259.1}{\mebi\byte} & {\bfseries \qty{0.029267}{\second}} & \qty{0.846}{\mebi\byte} \\
        64 & \qty{14.0}{\second} & \qty{69.3}{\mebi\byte} & \qty{110.4}{\second} & \qty{516.8}{\mebi\byte} & {\bfseries \qty{0.084140}{\second}} & \qty{3.055}{\mebi\byte} \\
        128 & \qty{54.3}{\second} & \qty{173.3}{\mebi\byte} & \qty{432.4}{\second} & \qty{1,008007812}{\gibi\byte} & {\bfseries \qty{0.291}{\second}} &  \qty{11.840}{\mebi\byte} \\
        256 & \qty{214.3}{\second} &  \qty{273.1}{\mebi\byte} & \qty{1711.5}{\second} & \qty{2,014550781}{\gibi\byte} & {\bfseries \qty{1.119}{\second}} & \qty{46.929}{\mebi\byte} \\
        \bottomrule
    \end{tabular}
\end{table}

\subsection{Performance of \otfpsiss and \otfpsissb}
\label{subsec:eval:secret_shared}

\cref{tab:eval:otfpsi_secret_shared_silent} shows the run time and communication cost of our secret-shared FPSI protocols, \otfpsiss and \otfpsissb. Compared to plaintext \otfpsi, \otfpsiss increases runtime around \qty{5}{\x} (on fast networks) and communication by only about $\qty{10}{\percent}$. Most of the additional run time is due to computation of the additional OTs which could be parallelized.

\otfpsissb reduces the number of OTs at the cost of additional communication. Over gigabit networking and for $\sizeA > 1$, \otfpsissb is about \qty{50}{\percent} faster than \otfpsiss while increasing communication by around \qty{2.5}{\x}. The benefit of \otfpsissb is more visible when instantiated with a more communication-heavy OT like SoftSpokenOT (Appendix, \cref{tab:eval:otfpsi_secret_shared_soft}). Here, \otfpsissb reduces communication by 
\qty{61}{\percent} and run time on slow networks by 
\qty{53}{\percent}.

\begin{table*}
    \centering
    \caption{Run time and communication of plaintext \otfpsi and secret-shared \otfpsiss and \otfpsissb with SilentOT for querier set size \sizeA and responder set size \sizeB (dimension $\dimension = 511$, threshold $\threshold = \floor{\dimension/16} = 31$). }
    \label{tab:eval:otfpsi_secret_shared_silent}
    \begin{tabular}{rrrrrrrrrrr}
        \toprule
         &  & \multicolumn{3}{c}{\otfpsi} & \multicolumn{3}{c}{\otfpsiss} & \multicolumn{3}{c}{\otfpsissb} \\
         \cmidrule(lr){3-5} \cmidrule(lr){6-8} \cmidrule(lr){9-11}
        \sizeA & \sizeB & Gigabit & Slow & Comm & Gigabit & Slow & Comm & Gigabit & Slow & Comm \\
        \midrule
        \num{64} & \num{64} & \qty{0.080}{\second} & \qty{1.005}{\second} & \qty{2.668}{\mebi\byte} & \qty{0.356}{\second} & \qty{1.371}{\second} & \qty{2.943}{\mebi\byte} & \qty{0.201}{\second} & \qty{1.240}{\second} & \qty{7.245}{\mebi\byte} \\
        \num[round-mode=none]{256} & \num[round-mode=none]{256} & \qty{1.046}{\second} & \qty{2.672}{\second} & \qty{40.745}{\mebi\byte} & \qty{5.388}{\second} & \qty{7.326}{\second} & \qty{44.761}{\mebi\byte} & \qty{2.648}{\second} & \qty{5.076}{\second} & \qty{113.768}{\mebi\byte} \\
        \num[round-mode=none,exponent-mode=input]{1024} & \num[round-mode=none,exponent-mode=input]{1024} & \qty{16.474}{\second} & \qty{29.570}{\second} & \qty{649.386}{\mebi\byte} & \qty{84.183}{\second} & \qty{101.471}{\second} & \qty{714.039}{\mebi\byte} & \qty{41.925}{\second} & \qty{86.795}{\second} & \qty{1.775}{\gibi\byte} \\
        \num[round-mode=none,exponent-mode=input]{4096} & \num[round-mode=none,exponent-mode=input]{4096} & \qty{266.900}{\second} & \qty{461.935}{\second} & \qty{10.143}{\gibi\byte} & \qty{1377.703}{\second} & \qty{1661.770}{\second} & \qty{11.155}{\gibi\byte} & \qty{674.221}{\second} & \qty{1459.881}{\second} & \qty{28.393}{\gibi\byte} \\
        \midrule
        \num{1} & \num[round-mode=none,exponent-mode=input]{16384} & \qty{0.261}{\second} & \qty{1.417}{\second} & \qty{10.257}{\mebi\byte} & \qty{1.369}{\second} & \qty{2.609}{\second} & \qty{11.317}{\mebi\byte} & \multicolumn{3}{c}{\multirow{3}{*}{\makecell{\emph{Batching not applicable}\\ \emph{for $\sizeA=1$}}}}  \\
        \num{1} & \num[round-mode=none,exponent-mode=input]{131072} & \qty{1.977}{\second} & \qty{4.705}{\second} & \qty{81.270}{\mebi\byte} & \qty{10.552}{\second} & \qty{13.305}{\second} & \qty{89.332}{\mebi\byte} &  &  &  \\
        \num{1} & \num[round-mode=none,exponent-mode=input]{524288} & \qty{7.812}{\second} & \qty{15.228}{\second} & \qty{324.704}{\mebi\byte} & \qty{42.093}{\second} & \qty{50.182}{\second} & \qty{356.729}{\mebi\byte} &  &  &  \\
        \bottomrule
    \end{tabular}
\end{table*}

\subsection{End-to-End Evaluation of \name}
\label{subsec:eval:e2e}
We evaluate the communication and computation cost of \name to show that it is practical and fulfills the requirements outlined in \S\ref{sec:syst:reqs}. Fulfilling \reqlinky{scalability}, we assume there are \num[round-mode=none,exponent-mode=input]{131072} existing registrations and \num[round-mode=none]{2048} new registrations. We assume $\dimension = 511$ and $\threshold = 132$ (see \S\ref{sec:embedding}).

\descr{Setup.}
Recall that during the setup phase, all field teams embed their existing records and send their secret shares to the two compute nodes. Computing the embedding is done locally and is relatively cheap: For \num[round-mode=none,exponent-mode=input]{131072} records, embedding can be done in \qty{388.2630}{\second} (see \cref{ap:emebdding}) and can be easily parallelized. 
Using pseudo-random secret sharing with a $256$-bit seed, a field team that submits \num[round-mode=none,exponent-mode=input]{131072} registrations has a total communication cost of
\qty{7,984405518}{\mebi\byte}. 

\descr{Offline Operation.} The field team embeds the \num[round-mode=none]{2048} new records which takes
\qty{6,066609375}{\second}. Sending secret-shared embeddings requires sending a total of
\qty{127,78125}{\kibi\byte} to the compute nodes. 
The nodes run a secret-shared FPSI protocol to compare the \num[round-mode=none]{2048} new registrations to the \num[round-mode=none,exponent-mode=input]{131072} existing registrations of other organizations. 
Using \otfpsiss with SilentOT, we estimate this takes a total of
\num{359,253}\;min over gigabit networking, requiring a communication of
\qty{178,5}{\gibi\byte}. \otfpsissb can reduce the run time to
\num{178,773333333}\;min with 
\qty{455,68}{\gibi\byte} communication. On average, even \otfpsissb utilizes only about a third of the available bandwidth, and hence both protocols could further benefit from parallelized computation.
Finally, the querier can retrieve the secret shares of the result 
(\qty{256,000244141}{\mebi\byte}) and recombine them. 

The offline operation mode (\reqlinky{offline}) of \name only requires very limited communication and computation by the querying organization (\reqlinky{local-efficient}), while no interaction is required by any other organization (\reqlinky{offline}). While the querier may need to wait multiple hours between submitting the new registrations and retrieving the results, there are typically no strict run time requirements for offline operation as long as the process can still happen, e.g., overnight. 

\descr{Online Operation.} Embedding a single record takes
\qty{2,962211609}{\milli\second} and its secret shares have a size of
\qty{96}{\byte}. The compute nodes can perform a query with \otfpsiss and SilentOT in \qty{10,6}{\second} with \qty{127}{\mega\byte} of communication. Using \otfpsiss with SoftSpokenOT can reduce the run time of a query to \qty{6.73}{\second} at the cost of increased communication (see Appendix, \cref{tab:eval:otfpsi_secret_shared_soft}). The querying organization can then retrieve the result shares which are
\qty{128,25}{\kilo\byte}. \name's online mode also requires very little resources from the querying field team (fulfilling \reqlinky{local-efficient}). It returns a result within seconds (\reqlinky{offline}), yet we acknowledge that a delay of \qtyrange{6.73}{10.6}{\second} might slow down registration processes. We expect that this is still practical, since deduplication can be interleaved with other steps of the registration process.

\subsection{Comparison to Related Work}
\label{subsec:eval:pprl}
\descr{\Mainzelliste \cite{Stammler2020}. } Our embedding-based approach provides comparable accuracy to Stammler \etal's \ac{smc} implementation \cite{Stammler2020} of the \EpiLink matching algorithm for \ac{pprl} (see \cref{ap:emebdding}). 
\Mainzelliste is prohibitively expensive for deduplication. Even if with a parallelized implementation and using fewer fields (as the authors), we estimate that \Mainzelliste would require
over $10$ days of computation
and \qty{154,298368}{\tebi\byte} of communication to deduplicate a batch in offline mode. %
\name with \otfpsissb can do this in only 
\qty{178.773}{\minute}, reducing total cost by a factor of 
\qty{84.1}{\x} and communication by 
\qty{347}{\x}.
We further estimate that an online query with \Mainzelliste would take a total of
\qty{440,40192}{\second}
(\qty{78.6432}{\second} online computation). In contrast, \otfpsi only takes \qty{10,6}{\second} in total.

\descr{Funshade \cite{Ibarrondo2023}. } The Funshade protocol allows threshold distance comparisons of vectors using $\Pi$-secret sharing and Function Secret Sharing (FSS).  
As such, it may seem to be more naturally suited for two non-colluding compute nodes.  
However, the authors only evaluate their protocol with a trusted third party (TTP) to generate $\Pi$-shares and FSS keys. 
Even with a TTP, we estimate that Funshade's setup phase for an online query would around \qty{30}{\second} over our slow network -- and even more when replacing the TTP with SMC. In contrast, \otfpsissb only needs \qty{13.3}{\second} total.

Lastly, since $\Pi$-shares embed Beaver triplets, in Funshade, they are re-created by the data holders for each comparison which does not work in our system model where data holders may be offline (violating \reqlinky{offline}) and putting load on the field teams (violating \reqlinky{local-efficient}).

Overall, Funshade is more expensive than \otfpsiss and \otfpsissb, and does not work in our system model. We provide a more detailed analysis in the full version. 

\section{Conclusion}
\label{sec:conclusion}
\noindent In this work, we proposed \name, a new privacy-preserving deduplication system for the humanitarian sector. We build on \otfpsi, a new FPSI protocol that outperforms all existing FPSI protocol without restrictive input assumptions.

\vspace{.1cm}
\descr{Acknowledgements.} Tim Rausch carried out this work as a member of the Saarbrücken Graduate School of Computer Science.

\section*{Ethics Considerations}\label{sec:ethics}

During the course of our research, no harm was caused. We did not incorporate human subjects into our research, nor did we gather any data about people. We deliberately worked with synthetic evaluation dataset.

We design a privacy-friendly deduplication system that guarantees strong 
privacy protection. 
Hence, it can be used in situations where non-private deduplication systems cannot and can offer assistance to more recipients.
We have carefully considered the impact of incorrectly being singled out as a duplicate, and have minimized the risk of this happening in the first place, and clearly positioned our system within a bigger system with additional checks and balances.

Yet, deduplication systems are not fully without a potential for harm, regardless of whether they are private or not.
The first harm is to those correctly identified as duplicates which may be outweighed by the fact that more people can receive aid. Secondly, malicious recipients could extract information if registration data is not verified (\S\ref{subsec:system:limitations}). Lastly, deduplication systems can be used for other means such as migration enforcement~\cite{reidy2017asylum}.
However, non-private deduplication systems already exists and are in use. These can already be misused, and our system does not increase the potential for harm with respect to existing systems.

We recognize that our construction could enable privacy-washing -- an inherent risk that can only thwarted by strong ethics considerations in its application.

\section*{LLM Usage Considerations}
An LLM-based tool (Grammarly) was used for editorial purposes in this manuscript, and all outputs were manually inspected and approved by the authors to ensure accuracy and originality.

\bibliographystyle{IEEEtranS}
\bibliography{99_references-dedup}

\appendices

\crefalias{section}{appendix}
\crefalias{subsection}{appendix}
\crefalias{subsubsection}{appendix}
\newcommand{\appendixsection}[1]{\section{#1}}

    \appendixsection{Oblivious Transfer}
    \label{ap:ot}
    \acreset{ot}

    \newcommand{\otbitlength}{\ell}
    1-out-of-$N$ \ac{ot} is a two party functionality between a querier \partyA and a responder \partyB that holds $N$ messages $m_0,\dots,m_{N-1} \in \bin^\otbitlength$ of length $\otbitlength$. \ac{ot} allows \partyA to learn $m_c$ for an arbitrary choice $c\in\mathbb{Z}_N$ such that (a) \partyB learns no information about \partyA's choice $c$ (nor the chosen message $m_c$), and (b) \partyA learns no information about the non-chosen messages $m_i$ for $i \in \mathbb{Z}_N \setminus \{c\}$. 

\begin{figure}
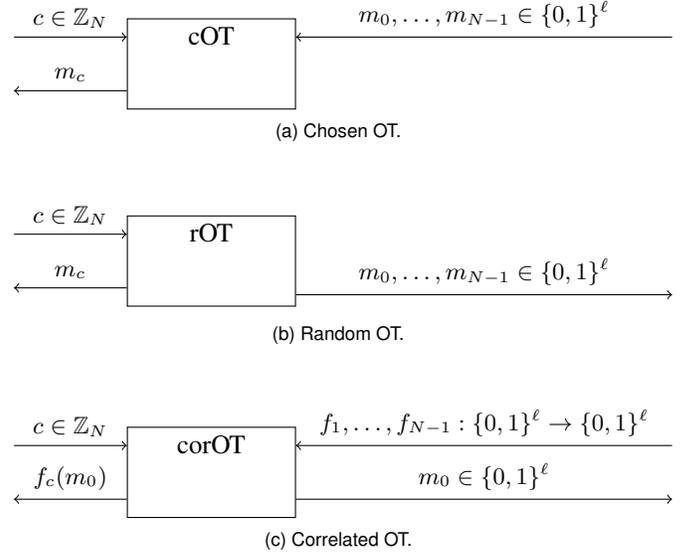

        \centering
        \subfloat[Chosen OT.]{
            \centering
            \label{fig:ap:ot:chosen}
            \begin{bbrenv}{A}
                \begin{bbrbox}[name=cOT,namepos=center,minheight=1.2cm]
                \end{bbrbox}
                \bbrmsgto{top=$c \in \mathbb{Z}_N$,length=1.5cm,topstyle={font=\small},fixedoffset=.25cm}
                \bbrqryfrom{top={$m_0,\dots,m_{N-1}\in\bin^\otbitlength$},length=5cm,topstyle={font=\small},fixedoffset=.25cm}
                \bbrmsgfrom{top={$m_c$},length=1.5cm,topstyle={font=\small},fixedboffset=.25cm}
            \end{bbrenv}
        }

        \vspace{1em}
        
        \subfloat[Random OT.]{
            \centering
            \label{fig:ap:ot:random}
            \begin{bbrenv}{A}
                \begin{bbrbox}[name=rOT,namepos=center,minheight=1.2cm]
                \end{bbrbox}
                \bbrmsgto{top=$c \in \mathbb{Z}_N$,length=1.5cm,topstyle={font=\small},fixedoffset=.25cm}
                \bbrqryto{top={$m_0,\dots,m_{N-1}\in\bin^\otbitlength$},length=5cm,islast=true,topstyle={font=\small},fixedoffset=.25cm}
                \bbrmsgfrom{top={$m_c$},length=1.5cm,topstyle={font=\small},fixedboffset=.25cm}
            \end{bbrenv}
        }

        \vspace{1em}

        \subfloat[Correlated OT.]{
            \centering
            \label{fig:ap:ot:correlated}
            \begin{bbrenv}{A}
                \begin{bbrbox}[name=corOT,namepos=center,minheight=1.2cm]
                \end{bbrbox}
                \bbrmsgto{top=$c \in \mathbb{Z}_N$,length=1.5cm,topstyle={font=\small},fixedoffset=.25cm}
                \bbrqryfrom{top={$f_1,\dots,f_{N-1}: \bin^\otbitlength \rightarrow \bin^\otbitlength$},length=5cm,topstyle={font=\small},fixedoffset=.25cm}
                \bbrqryto{top={$m_0\in\bin^\otbitlength$},length=5cm,topstyle={font=\small},fixedboffset=.25cm}
                \bbrmsgfrom{top={$f_c(m_0)$},length=1.5cm,topstyle={font=\small},fixedboffset=.25cm}
            \end{bbrenv}    
        }
        
        \caption{Overview over 1-out-of-$N$ OT variants.}
        \label{fig:ap:ot}
    \end{figure}

    \descr{Variants.} \ac{ot} functionalities can be classified by how much control the sender has over the messages. In \emph{chosen OT} (\cref{fig:ap:ot:chosen}), the sender can input a set of arbitrarily chosen messages. In \emph{random OT} (\cref{fig:ap:ot:random}) the messages are randomly sampled by the protocol and then output to the sender, giving the sender no control over the messages.

    In \emph{correlated OT} (\cref{fig:ap:ot:correlated}), only one message $m_0$ is randomly chosen by the protocol. The remaining messages $m_1, \dots, m_{N-1}$ are computed by evaluating arbitrary correlation functions $f_1,\dots,f_{N-1}$ chosen by \partyB on $m_0$. The querier \partyA learns $m_c = f_c(m_0)$ where $f_0$ is the identity function.

    \descr{Implementations.} Direct OT implementations typically rely on public-key techniques \cite{Naor2001,Chou2015}. \acf{OTe} protocols \cite{Ishai2003,Kolesnikov2016,Roy2022,Boyle2019} can efficiently extend OTs -- i.e., perform a large number of OTs given a few base OTs and typically using only symmetric key techniques. OTe protocols usually only provide random OT functionality \cite{Ishai2003,Roy2022}.

    \descr{Constructions. } Random OT can be transformed into chosen and correlated OT at the cost of additional communication. In both cases, \partyA and \partyB run a random OT with choice $c$, returning the random messages $\omega_0,\dots,\omega_{N-1}$ to \partyB and $\omega_c$ to \partyA. For chosen OT, \partyB uses these random messages to mask the chosen messages as $\mu_i = m_i \xor \omega_i$ and sends $\mu_0, \dots, \mu_{N-1}$ to \partyA, who can only reconstruct $m_c = \mu_c \xor \omega_c$. For correlated OT, \partyB uses $m_0 = \omega_0$, computes $\mu_i = f_i(m_0) \xor \omega_i$, and sends $\mu_1,\dots,\mu_{N-1}$ to \partyA, who can only reconstruct $m_c = \mu_c \xor \omega_c$ (with $\mu_0 = 0^\otbitlength$). The construction for correlated OT requires one fewer message to be sent, making correlated OT from random OT (like OTe) more efficient that chosen OT from random OT.

    \descr{Large Messages.} For large $\otbitlength$-bit messages, a random OT can also be implemented by executing a random OT for \secpar-bit messages (where \secpar{} is a security parameter) and then using a public pseudo-random function $F: {\bin}^\secpar \rightarrow {\bin}^\otbitlength$ to extend the random \secpar-bit messages into pseudo-random $\otbitlength$-bit messages. Combining this with the construction above provides 1-out-of-$N$ chosen and correlated OT for $\otbitlength$-bit messages at the cost of one \secpar-bit random OT, $N$ evaluations of $F$, and $N\otbitlength$ bits of communication for chosen OT ($(N-1)\otbitlength$ bits for correlated OT). 

\appendixsection{Full Proofs}
\label{ap:proofs}

\descr{\otfpsi. }
\otfpsi (\cref{fig:fpsi:otfpsi}) securely implements \ffpsi (\cref{fig:sytem:fpsi_ideal}) against semi-honest adversaries in the OT-hybrid model: %
Let \otfpsihyb denote the hybrid protocol with an idealized OT.

\subdescr{Correctness. } Let $\ffpsi(\setA, \setB) = (\ffpsipart{\partyA}(\setA, \setB), \bot)$ be the ideal FPSI functionality (\cref{fig:sytem:fpsi_ideal}). We denote the output of an execution of \otfpsihyb by $\protooutput{\otfpsihyb}(\setA, \setB) = (\protooutputpart{\otfpsihyb}{\partyA}(\setA, \setB), \bot)$.

As for the single comparison (\S\ref{subsec:fpsi:single_comparison}), $\vv{D_i}[k] - \vv{M_i}[k] \bmod \modulus = \hammingdist(q_i, r_k)$ holds for $i \in \range{\sizeA}, k \in \range{\sizeB}$. Hence, 
\begin{equation*} 
    \begin{split}
        (i,k) \in  \protooutputpart{\otfpsi}{\partyA}(\setA, \setB) &\iff b_{i,k} = 1 \\
        &\iff \vv{D_i}[k] - \vv{M_i}[k] \bmod \modulus \leq \threshold \\
        &\iff \hammingdist(q_i, r_k) \leq \threshold\\
        &\iff (i,k) \in \ffpsipart{\partyA}(\setA, \setB).\\
    \end{split}
\end{equation*}
Thus, $\protooutput{\otfpsihyb}(\setA, \setB) = \ffpsi(\setA, \setB)$ and \otfpsi is correct. 

\subdescr{Security. } A party $P$'s view $\view{\otfpsihyb}{P}(\setA, \setB)$ consists of its input and the messages it receives during the protocol execution.
To prove security, we show that a party's view of \otfpsihyb can be simulated based on inputs and outputs of the idealized functionality $\mathcal{F}_\text{FPSI}$, and simulated views are indistinguishable from views in the hybrid protocol.

\subdescr{Simulating \partyB's View. } During the execution of \otfpsihyb, \partyB receives no messages. Its view consists only of its input and can be trivially simulated by $\simulator_\partyB(\setB, \bot)$ that only outputs \setB. 
The simulated and hybrid views follow the same distribution, i.e., $\{\simulator_\partyB(\setB, \bot)\}_{\setA,\setB} \disteq \{\view{\otfpsihyb}{\partyB}(\setA,\setB)\}_{\setA,\setB}$.

\subdescr{Simulating \partyA's View. } During the execution of \otfpsihyb, \partyA receives $\vv{m_{0,0}}, \dots, \vv{m_{\sizeA,\dimension}}$ and $b_{0,0}, \dots, b_{\sizeA,\sizeB}$. We construct a simulator $\simulator_\partyA(\setA, I)$ which samples $\vv{m_{0,0}}, \dots, \vv{m_{\sizeA,\dimension}} \sample \group^\sizeB$, sets $b_{i,k} = ((i,k) \in I)$ for $i \in \range{\sizeA}, k \in \range{\sizeB}$, and outputs the messages in the correct order.
In \otfpsihyb, the $\vv{m_{0,0}}, \dots, \vv{m_{\sizeA,\dimension}}$ are uniformly random in $\group$ and the $b_{i,k}$ are an immediate encoding of the result $\ffpsipart{\partyA}(\setA, \setB)$. Hence, simulated and hybrid views follow the same distribution: $\{\simulator_\partyA(\setA, \ffpsipart{\partyA}(\setA, \setB))\}_{\setA,\setB} \disteq \{\view{\otfpsihyb}{\partyA}(\setA,\setB)\}_{\setA,\setB}$.

Therefore, \otfpsi securely implements \ffpsi. %

\descr{\otfpsiss. } \otfpsiss (\cref{fig:fpsi:otfpsiss}) securely implements \fssfpsi in the OT-hybrid model against semi-honest adversaries.
Let \otfpsisshyb denote the hybrid protocol with idealized OT.

\subdescr{Correctness. } We denote the probabilistic \fssfpsi functionality by $\fssfpsi(x, y) = (\fssfpsipart{1}(x, y), \fssfpsipart{2}(x, y))$  with $x = (\shareA{\setA}, \shareA{\setB})$ and $y = (\shareB{\setA}, \shareB{\setB})$. For $(\shareA{P}, \shareB{P}) \gets \protooutput{\otfpsiss}(x, y)$:
\begin{equation*} 
    \begin{split}
        \shareA{P}_{i,j} &= (D_{i,j} - M_{i,j} \bmod \modulus \leq \threshold) \xor \shareB{P}_{i,j} \\
        &= (\hammingdist(\shareA{\elementA_i} \xor \shareA{\elementB_j}, \shareB{\elementA_i} \xor \shareB{\elementB_j}) \leq \threshold) \xor \shareB{P}_{i,j} \\
        &= (\hammingdist(\elementA_i, \elementB_j) \leq \threshold) \xor \shareB{P}_{i,j}. \\
    \end{split}
\end{equation*}
Hence, $\protooutput{\otfpsiss}(x,y) = ((\hammingdist(\elementA_i, \elementB_j) \leq \threshold)_{i,j} \xor \shareB{P}, \shareB{P})$ 
where $\shareB{P}$ is uniformly random. By definition, $\fssfpsi(x, y) = ((\hammingdist(\elementA_i, \elementB_j) \leq \threshold)_{i,j} \xor \shareB{M}, \shareB{M})$ with uniformly random $\shareB{M}$. Hence, $\{\fssfpsi(x,y)\}_{x,y} \disteq\{\protooutput{\otfpsiss}(x,y)\}_{x,y}$ holds.

\subdescr{Simulating \serverB's View. } In the hybrid protocol \otfpsisshyb, \serverB receives no messages. As for \partyB in \otfpsi, a simulator $\simulator_2(y, \bot)$ that outputs only \serverB's input $y$ perfectly simulates \serverB's view, i.e., 
 $\{\simulator_2(y, \bot)\}_{x,y} \disteq \{\view{\otfpsisshyb}{2}(x,y)\}_{x,y}$. Hence, it follows that $\{\simulator_2(y, \bot), \fssfpsi(x,y)\}_{x,y} \disteq \{\view{\otfpsisshyb}{2}(x,y), \protooutput{\otfpsisshyb}(x,y)\}_{x,y}$.

\subdescr{Simulating \serverA's view. } In the hybrid \otfpsisshyb, \serverA receives the messages $m_{1,1,1}, \dots, m_{\sizeA,\sizeB,\dimension}$ and $\shareA{P}_{1,1}, \dots, \shareA{P}_{\sizeA, \sizeB}$. We construct a simulator $\simulator_1(x, \shareA{M})$ which samples $m_{1,1,1},$
$\dots, m_{\sizeA,\sizeB,\dimension} \sample \group$, samples $\shareA{P}_{1,1}, \dots, \shareA{P}_{\sizeA, \sizeB} \sample \bin$, and outputs the messages in the correct order.

As all messages received by \serverA in \otfpsisshyb are uniformly random, simulated views have the same distribution $\{\simulator_1(x_1, \fssfpsi^1(x, y))\}_{x,y} \disteq \{ \view{\otfpsisshyb}{1}(x,y) \}_{x, y}$. It follows that $\{\simulator_1(x, \fssfpsi^1(x, y)), \fssfpsi(x,y)\}_{x,y} \disteq \{ \view{\otfpsisshyb}{1}(x,y), \protooutput{\otfpsisshyb}(x,y) \}_{x, y}$.

We conclude that \otfpsiss securely implements \fssfpsi. %

\descr{\otfpsissb. } \otfpsissb (\cref{fig:fpsi:otfpsissb}) securely implements \fssfpsi in the OT-hybrid model against semi-honest adversaries.
Let \otfpsissbhyb denote the hybrid protocol with idealized OT.

\subdescr{Correctness. } With $i \in \range{\sizeA}, j \in \range{\sizeB}, j \in \range{\dimension}, z = (i,j,k)$ it holds that
\begin{equation*} 
    \begin{split}
        d_z &= \prfff{c_z}{X_{i,k}} + \prfff{c_z}{Y_{j,k}} + \mu_{z,c_z} \bmod \modulus\\
            &= \prfff{c_z}{X_{i,k}} + \prfff{c_z}{Y_{j,k}} + m_{z,c_z} \\ 
            & \quad - \prfff{c_z}{X_{i,k}^{\shareA{\elementA_i}[k]}} - \prfff{c_z}{Y_{j,k}^{\shareA{\elementB_j}[k]}} \bmod \modulus \\
            &= m_{z,c_z} = m_{z,2\shareA{\elementA_i}[k] + \shareA{\elementB_j}[k]}.
    \end{split}
\end{equation*}
By construction, we have $m_{z,2\shareA{\elementA_i}[k] + \shareA{\elementB_j}[k]} = m_z + (\shareA{\elementA_i}[k] \xor \shareA{\elementB_j}[k] \xor \shareB{b}_z)$. Hence, $d_z - m_z \bmod \modulus = \shareA{\elementA_i}[k] \xor \shareA{\elementB_j}[k] \xor \shareB{\elementA_i}[k] \xor \shareB{\elementB_j}[k] = \elementA_i[k] \xor \elementB_j[j]$. Since $D_{i,j} = \sum_{k=1}^{\dimension} d_{(i,j,k)} \bmod p$ and $M_{i,j} = \sum_{k=1}^{\dimension} m_{(i,j,k)} \bmod p$, it follows that $D_{i,j} - M_{i,j} \bmod \modulus = \hammingdist(\elementA_i, \elementB_j)$.

The threshold comparison step of \otfpsissb is identical to \otfpsiss. Using the same argument, it follows that the outputs of \otfpsissb and \fssfpsi follow the same distribution: $\{\fssfpsi(x_1,x_2)\}_{x_1,x_2} \disteq\{\protooutput{\otfpsissb}(x_1,x_2)\}_{x_1,x_2}$.

\subdescr{Simulating \serverB's View. } As in \otfpsisshyb, \serverB's view can be perfectly simulated since \serverB receives no messages.

\subdescr{Simulating \serverA's view. } \serverA receives the messages $X_{1,1},$ 
$\dots, X_{\sizeA, \dimension}$, $Y_{1,1}, \dots, Y_{\sizeB, \dimension}$, $\mu_{(1,1,1),1}, \dots,\mu_{(\sizeA,\sizeB,\dimension),3}$ and $\shareA{P}_{1,1}, \dots, \shareA{P}_{\sizeA, \sizeB}$. We construct $\simulator_1(x_1, \shareA{M})$ that samples these messages uniformly random from their respective domain and outputs them in the correct order.
Views generated by the simulator are computationally indistinguishable (c.id.) from views in the hybrid protocol: We first argue that, for a single $z = (i,j,k)$, the $\mu_{z,1}, \mu_{z,2}, \mu_{z,3}$ in the hybrid view are c.id. from random. In the hybrid protocol, the ideal OT perfectly hides one of the random PRF keys $X^0_{i,k}$ and $X^1_{i,k}$ (and one of $Y^0_{j,k}$ and $Y^1_{j,k}$) from \serverA. %

If $c_z = 0$, all the $\mu_{c_z,x}$ contain at least one term that is an evaluation of the PRF $F$ with a random key that is not contained in \serverA's view. Since \serverB does not evaluate $F$ twice on the same input under the same key, these terms are c.id. from random by the PRF property of $F$. With this, the  $\mu_{z,1}, \mu_{z,2}, \mu_{z,3}$  in the view are also c.id. from random.

If $c_z \neq 0$, the $\mu_{z,x}$ for $x \neq c_z$ are c.id. from random following the same argument. Hence, they computationally hide information about $m_{z,0}$ and $m_{z,1}$. 
The value $\mu_{z,c_z}$ is a sum which contains $m_{z,c_z}$ which in turn is a sum containing at least one PRF evaluation with a key not included in the view, rendering $m_{z,c_z}$ c.id. from random. Thus, $\mu_{z,c_z}$ is also c.id. from random.

For one $z$, the $\mu_{z,x}$ in \serverA's hybrid view can, hence, be replaced with random values while keeping the views c.id. Therefore, hybrid views are c.id. from intermediate hybrid views where the $\mu_{z,x}$ are replaced by random values for all $z$. Since all messages in these intermediate views have the same distribution as views generated by the simulator. Hence, %
$\{\simulator_1(x_1, \fssfpsi^1(x_1, x_2))\}_{x_1,x_2} \distindist \{ \view{\otfpsissbhyb}{1}(x_1,x_2) \}_{x_1, x_2}$, and, as a result, it holds that $\{\simulator_1(x_1, \fssfpsi^1(x_1, x_2)), \fssfpsi(x_1,x_2)\}_{x_1,x_2} \distindist \{ \view{\otfpsissbhyb}{1}(x_1,x_2), \protooutput{\otfpsissbhyb}(x_1,x_2) \}_{x_1, x_2}$

Hence, \otfpsissb securely implements \fssfpsi. %

    \appendixsection{Embeddings and Matching}
    \subsection{Details about our Embedding}

\label{ap:emebdding}
    We present our embedding strategy in more detail. 
    Our survey of the literature did not highlight an embedding into Hamming space proposed in the record linkage domain that would be suitable for humanitarian organizations. Thus, we propose a new embedding \embedding and evaluate it to show that it works well for humanitarian deduplication. That being said, we stress that \name is agnostic to the concrete embedding used, and this construction may be easily replaced.

\descr{Construction. }
To transform a record into a bit string, we first compute its set of \qgrams (i.e., all substrings of length \gramlen). We then construct the embedding by concatenating 1-bit locality-sensitive hash (LSH) values of the \qgrams set. As \ac{lsh} preserves the similarity of the input (i.e., the set of \qgrams), the number of identical bits in embedded strings can be used to estimate the similarity of the sets of \qgrams, and therefore the similarity of the original records.

\subdescr{Embedding into Hamming Space. } A locality-sensitive hashing scheme \cite{Charikar2002} is a family of functions $\mathcal{H}$ for a similarity function \attributesim{}, such that $\probsub{h \sample \mathcal{F}}{h(x) = h(y)} = \attributesim{}(x, y)$. Charikar \cite{Charikar2002} proposes to embed a similarity with an LSH with range $\bin$ into Hamming space by concatenating \dimension individual LSH values. The more similar two objects are, the more individual bit LSH values will match, and the lower the Hamming distance will be. We apply this construction to the Minhash \cite{Broder2000} LSH for the Jaccard similarity of sets. 

\subdescr{Handling Records.}
To apply Minhash to the structured registration records of humanitarian organizations, we first need to convert these records to sets. We do this by computing the \qgrams of each attribute individually and then uniting these sets while keeping \qgrams of different attributes domain-separated.
Assuming a record $x$ consists of the attributes $x[1],\dots,x[\attributecount]$, we convert $x$ into the set
\begin{equation}
    X = \bigcup {}_{i \in \range{\attributecount}} \{(i \concat g) \mid g \in \grams{\gramlen}{x[i]}\}
    \label{eq:hamming_transformation:set}
\end{equation}
and then transform the set $X$ to a bit string of length \dimension using the Minhash-based embedding described above.

\descr{Evaluation. }
    As our work deals with recipients' personal information, we do not work with real humanitarian data but with public, partially synthetic datasets. 
    For a meaningful evaluation, we need a dataset with ground truth relevant for our use case. Prior works often used standard bibliographic, location, or e-commerce datasets \cite{Koepcke2010, He2017} whose attributes and duplicates are not representative of those encountered by humanitarian organizations.
    We curate a new dataset by extending the public North Carolina Voter Registration database~\cite{NCVR}, enabling us to imitate real duplicates in a controlled manner with a determined ground truth. We detail the preprocessing and synthetic duplicate generation in \cref{ap:dataset}. 
    To match humanitarian use cases, the dataset consists of the following fields: first and last names, date of birth, gender, mother's first and last name, and father's first name.

    To illustrate the accuracy of our embedding, we compare it to the \EpiLink matching algorithm \cite{Contiero2005} that has also been used to implement \ac{pprl} \cite{Stammler2020}. We expand on the parameter choices for \EpiLink in \cref{ap:params}. We also compare our embedding to the Jaccard similarity on domain-separated \qgrams as defined in \cref{eq:hamming_transformation:set}.

To compare all three methods, we sample a reference database of \num[round-mode=none,exponent-mode=input]{131072} records and a test dataset consisting of both test records that have a duplicate in the reference database and test records that do not. For each test record, we compute the similarity to all records in the reference database and determine the maximum similarity. We classify each test record as a duplicate or non-duplicate by comparing the maximum similarity to a range of thresholds.

\begin{figure}
    \centering
    \subfloat[ROC curve for embedding, \EpiLink, and Jaccard baselines.]{
        \label{fig:hamming_transformation:eval_roc}
        \includegraphics[width=0.45\columnwidth]{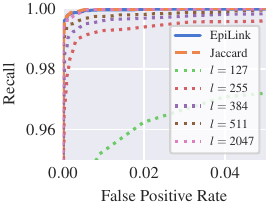}
    }
    \hfill
    \subfloat[FPR and FNR over normalized threshold $\threshold / \dimension$ for $\dimension = 511$.]{
        \label{fig:hamming_transformation:eval_fpr_fnr}
        \includegraphics[width=0.45\columnwidth]{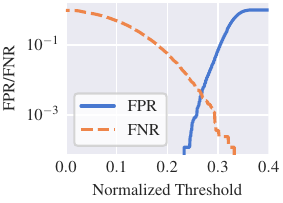}
    }
    \caption{Accuracy of classifying records as duplicate/non-duplicate in a database of \num[round-mode=none,exponent-mode=input]{131072} records ($\gramlen = 2$).}
    \label{fig:hamming_transformation:eval}
\end{figure}

\cref{fig:hamming_transformation:eval_roc} shows that \EpiLink, the Jaccard similarity, and our embedding with $\dimension=2047$ essentially provide the same accuracy. It is only insignificantly worse for $\dimension=511$, but accuracy measurably degrades for smaller \dimension. 

\subdescr{Optimal Threshold.} To use the embedding with \name, we need to set a Hamming distance threshold \threshold. To achieve the target false positive rate (FPR) of less than \qty{0.1}{\percent} (\reqlinky{fpr}) while keeping the false negative rate (FNR) low, we choose the maximum Hamming distance threshold satisfying this constraint. For $\dimension = 511$, this point is at $\threshold = 132$ (see \cref{fig:hamming_transformation:eval_fpr_fnr}). For this threshold, we achieve an FPR of \qty{000.09765625}{\percent} and an FNR of \qty{000,57373046875}{\percent}. For lower \dimension, we observe a higher FNR at the same FPR: \qty{001,2451171875}{\percent} for $\dimension = 384$ ($\threshold = 95$), \qty{002,23388671875}{\percent} for $\dimension = 255$ ($\threshold = 50$), and \qty{010,94970703125}{\percent} for $\dimension=127$ ($\threshold = 24$).

Our embedding is able to achieve the target false-positive rate (\reqlinky{fpr}) while minimizing false negatives at $\threshold \approx \dimension/4$.

\subdescr{Lower Thresholds.} Existing FPSI protocols typically use a small threshold relative to the dimension. To evaluate how our embedding performs in such a situation, we report the false-negative rates at select thresholds in \cref{tab:embedding:smaller_thresh}.

\subdescr{Performance.}
    Computing this embedding is done locally and offline by the party holding the registration records. 
    The computation is cheap and can be easily parallelized for multiple records. 
    Using a naive Python implementation, we can compute the embedding of one record for $\dimension = 511$ in 
    \qty{2,962211609}{\milli\second}.

\descr{Take-away.} Our embedding works well for deduplication as it provides high accuracy and can be computed efficiently. However, we stress that \name is agnostic to the underlying embedding, given that each party can compute the embeddings of its records locally.  %
    Should this specific transformation not translate well to other applications, potential alternatives include working on subsets of attributes \cite{Han2025}, weighted versions of Minhash \cite{Wu2022,Adir2022}, SimHash \cite{Manku2007}, or approaches based on machine learning \cite{Kerschbaum2023}.

    \begin{table}[t]
        \centering
        \caption{False-negative rates of the embedding with smaller thresholds. Empty cells: FPR exceeds \qty{.1}{\percent}.}
        \label{tab:embedding:smaller_thresh}
        \begin{tabular}{@{}r@{\hspace{2mm}}r@{\hspace{2mm}}r@{\hspace{2mm}}r@{\hspace{2mm}}r@{\hspace{2mm}}r@{}}
            \toprule
            & \multicolumn{5}{c}{\threshold} \\
            \cmidrule(l{0mm}r{0mm}){2-6}
            \dimension & 4 & 8 & 16 & 32 & 64 \\
            \midrule
            511 & \qty{096.5087890625}{\percent} & \qty{096.00830078125}{\percent} & \qty{090.4052734375}{\percent} & \qty{073.6328125}{\percent} & \qty{034.130859375}{\percent} \\
            255 & \qty{095.78857421875}{\percent} & \qty{090.61279296875}{\percent} & \qty{073.47412109375}{\percent} & \qty{035.36376953125}{\percent} & \\
            127 & \qty{090.1123046875}{\percent} & \qty{073.03466796875}{\percent} & \qty{036.04736328125}{\percent} & & \\
            \bottomrule
        \end{tabular}
    \end{table}

    \subsection{Embeddings into Euclidean Space}
    \label{ap:rw_euclidean}

    In the following, we present embeddings of records into Euclidean space that were proposed in the record linkage domain and estimate the embedding dimension necessary in the context of humanitarian deduplication.  We again assume that records consist of five string fields, a date of birth and a gender attribute.

    Li \etal~\cite{Li2006} present StringMap, an algorithm to transform strings into vectors in Euclidean space for non-private record linkage. They embed each attribute individually and recommend a dimension between 15 and 25 for each attribute. For our records consisting of five string fields (neglecting date of birth and gender), this would result in a total dimension of at least 75 to 125. However, their embedding algorithm operates on the database of all records in a non-private setting. It is unclear how this algorithm could be used in a private setting where records are distributed among multiple parties. 

    Bonomi \etal~\cite{Bonomi2012} propose to transform records into vectors in Euclidean space based on the occurrence of frequent grams in the record. The frequent grams are extracted from the entire database using differentially-private methods.
    In their evaluation, they chose a dimension of 75 for one single attribute. Applying this approach to records with more attributes would likely require an even higher dimension.

    Scannapieco \etal~\cite{Scannapieco2007} propose using SparseMap, a Lipschitz embedding that transforms records into vectors in Euclidean space by expressing them in terms of their similarity to a set of reference values.
    Their evaluation does not allow us to confidently predict the accuracy of their embedding in our setting. Still, we estimate that to achieve sufficient accuracy when comparing query records to large databases of records, we need at least a dimension of 30 for their evaluation data. As their data consists only of first and last name, we estimate we would need at least a dimension of 60 to embed the five string attributes of our records.

    \subsection{Synthetic Deduplication Dataset}
    \label{ap:dataset}

    We base our dataset on the North Carolina Voter Registration database (NCVR) which contains all registered voters in the U.S. state of North Carolina \cite{NCVR}. %

    \descr{Pre-Processing} We pre-process the database as follows: We remove all records that miss a first name, last name, or year of birth. 
    We extend the year of birth to a full date of birth by randomly sampling both day and month. For each record, we add the father's first name and mother's first and last name by randomly sampling a male and female first name and a last name from the database.  

    \descr{Synthetic Duplicate Generation.} We synthetically generate duplicates by applying perturbations to existing records. 
    For each generated duplicate, we apply up to four perturbations to randomly chosen attributes. Each perturbation is chosen at random from an attribute-specific list. 
    With a probability of $1/16$ for each perturbation, a destructive modification is chosen (like deleting the value or replacing it with a random value). 
    Otherwise, a non-destructive perturbation is applied, e.g., inserting, deleting, replacing, or swapping characters. 
    For fixed-length fields, we do not apply length-changing perturbations. 
    For the date of birth, we also include a perturbation that sets it to January 1st to emulate that exact dates of birth may not always be available in humanitarian contexts. When the gender field is selected for perturbation, its value is randomly replaced.

    \subsection{\EpiLink Parameter Choices}
    \label{ap:params}
    To evaluate the \EpiLink matching functionality \cite{AlLawati2005,Stammler2020}, we need to choose fields, weights, and attribute similarity metrics.
    We choose the same comparison mechanisms and parameters as Stammler \etal~\cite{Stammler2020}: 
    fields are converted into $2$-grams and then inserted into Bloom Filters of length \qty{500}{\bit} using \num{15} hash functions. Additionally, we split the date of birth into its individual components. 

    As suggested by Contiero \etal~\cite{Contiero2005}, we determine an attribute's weight based on its average frequency of values $f_i$ and its error rate $e_i$ as:
    \[
        w_i = \log_2\left(\frac{1-e_i}{f_i}\right).
    \]
    We determine attribute frequencies and error rates based on the parameter selection dataset: We sample $2^{18}$ original records and compute the average frequency of values for each attribute. To determine the error rates, we compare each of the sampled original records to its synthetically generated duplicate and for each attribute determine the error rate. A record-pairs has an error in an attribute if the attribute's values are not exactly equal.
    We show the attributes for deduplication and their associated weights in \cref{tab:linkage_attributes}. %

    \begin{table}
        \centering
        \small
        \caption{Overview of weights used for deduplication with attribute-level similarity metrics, average frequencies $f_i$, error rates $e_i$, and weights $w_i$. \attributesimds denotes fuzzy comparison using Bloom Filters, \attributesimeq denotes comparison by equality.}
        \label{tab:linkage_attributes}
        \begin{tabular}{llrrr}
            \toprule
            Attribute & Metric & $f_i$& $e_i$ & $w_i$ \\
            \midrule
            \texttt{first\_name} & \attributesimds & \num{3.5459735470373385e-05} & \num{0.30162284650481436} & \num{9.888116729662812}\\
            \texttt{last\_name} & \attributesimds & \num{1.766004415011037e-05} & \num{0.301724631212735} & \num{10.585064119923045}\\
            \texttt{gender} & \attributesimeq & \num{0.5} & \num{0.15364456176757812} & \num{0.5263313127046735} \\
            \texttt{dob\_year} & \attributesimeq & \num{0.01} & \num{0.13534927368164062} & \num{4.459740547791827} \\
            \texttt{dob\_month} & \attributesimeq & \num{0.0833333333} & \num{0.19705963134765625} & \num{2.2654318212882427} \\
            \texttt{dob\_day} & \attributesimeq & \num{0.0333333333} & \num{0.20445632934570312} & \num{3.1724678460652065} \\
            \texttt{first\_n\_mother} & \attributesimds & \num{3.480076561684357e-05} & \num{0.3032583939600874} & \num{9.90453051094372} \\
            \texttt{last\_n\_mother} & \attributesimds & \num{1.76678445229682e-05} & \num{0.3032722473144531} & \num{10.58240372093662} \\
            \texttt{first\_n\_father} & \attributesimds & \num{5.5266939316900624e-05} & \num{0.30357095173245136} & \num{9.441546310671814} \\
            \bottomrule
        \end{tabular}
    \end{table}

    \subsection{Matching with Private Approximate Jaccard} %
    \label{ap:jacard_lsh}

    Adir \etal~\cite{Adir2022} present a different approach for matching that does not rely on \ac{smc}: Using Minhash \cite{Broder2000}, they transform each record to a number of fingerprints such that, for a pair of records with high Jaccard similarity, one or more fingerprints will be equal with high probability. This transformation can be used for private matching by running \ac{psi} on these fingerprints.
    Han \etal~\cite{Han2025} extend this to angular similarity and provide a more detailed analysis of parameter choices.

    The main strength of this approach is that it overcomes the need for explicit pairwise matching as it allows fuzzily comparing sets of records using \ac{psi} in sub-quadratic run time.

    \descr{Construction.} Adir \etal~\cite{Adir2022} propose to generate \lshb{} fingerprints for each record using the Minhash LSH scheme \cite{Broder2000}. Each fingerprint consists of \lshr{} Minhash values of the record. Specifically, we convert a record $x$ into a set $X$ by computing its \qgrams or domain-separated \qgrams as in \cref{eq:hamming_transformation:set}. For permutations $\pi_0, \dots, \pi_{\lshb \cdot \lshr - 1}$, the fingerprints of $X$ are defined as
    \[
        f_i(X) = (m_{\pi_{i \cdot \lshr}}(X), \dots, m_{\pi_{i \cdot \lshr + \lshr - 1}}(X)).
    \]
    where $m_\pi$ is a Minhash function.
    For sets $X$ and $Y$, the probability that one of the fingerprints collides depends on their similarity:
    \[
        \Pr[\exists i < \lshb: f_i(X) = f_i(Y)] = 1 - (1 - \simjaccard(X,Y)^\lshr)^\lshb.
    \]
    Han \etal~\cite{Han2025} analyze how to choose parameters for this construction such that a collision likely occurs if the similarity is greater than a given threshold \threshold.

    \descr{Evaluation. }
    As for our embedding, we use domain-separated \qgrams (\cref{eq:hamming_transformation:set}) with $\gramlen = 2$. We compress the individual fingerprint vectors $f_i(x)$ into one value using a hash function.
    Han \etal~\cite{Han2025} choose the number of fingerprints as $\lshb \in \{30,50,100\}$. %
    For each $\lshb$, we evaluate a range of values for $\lshr$ and present the results as a ROC curve.

    \begin{figure}
        \centering
        \includegraphics{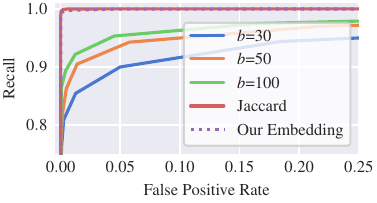}
        \caption{ROC curve for matching with Private Approximate Jaccard \cite{Adir2022} for different number of fingerprints \lshb{}, exact Jaccard, and our embedding ($\dimension = 511$) for $\sizeB = 2^{17}$.}
        \label{fig:jacard_lsh_roc_131k}
    \end{figure}

    \descr{Results.} \cref{fig:jacard_lsh_roc_131k} compares the accuracy of this private approximation of the Jaccard similarity, the exact Jaccard similarity, and our embedding (\cref{ap:emebdding}) when matching one record against a database of $\sizeB = \num[exponent-mode=input,round-mode=none]{131072}$ records. %

    We observe that the private Jaccard approximation does not achieve an accuracy that is comparable to the Jaccard similarity or our embedding. To achieve our target FPR of less than \qty{0.1}{\percent} (\reqlinky{fpr}), we would need to accept a recall of only \qty{086.46240234375}{\percent} for $b = 100$ (and only \qty{076.123046875}{\percent} for $b=30$), while our embedding is able to achieve a significantly higher recall of 
    \qty{99,426269531}{\percent} (see \cref{ap:emebdding}).

    \appendixsection{Evaluation}

    \subsection{Evaluation with SoftSpokenOT}\label{ap:eval:soft}

    As discussed in \S\ref{sec:evaluation}, we used SilentOT as a building block in our construction. Yet, \otfpsi can be instantiated with any OT block. Thus, for completeness, we also evaluate \otfpsi with SoftSpokenOT~\cite{Roy2022}.  

    SilentOT and SoftSpokenOT provide a different trade-off in terms of communication and computation cost: While SilentOT is significantly more communication-efficient, it requires more computation and different cryptographic hardness assumptions. 

\begin{figure}
        \centering
        \includegraphics[width=0.5\columnwidth]{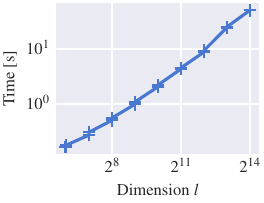}
        \caption{Run time of \otfpsi for $\sizeA=\sizeB=256$ by dimension \dimension ($\threshold=42$, Gigabit, SoftSpoken OTe).}
        \label{fig:eval:otfpsi_dimension}
    \end{figure}

    In \cref{fig:eval:otfpsi_dimension}, we confirm the theoretical relationship between \otfpsi's runtime and the dimension $\dimension$ (i.e., $\bigO{\dimension \log \dimension}$). 
    In \cref{tab:eval:otfpsi_performance_softspoken}, we present the overall performance of \otfpsi using SoftSpokenOT. \cref{tab:eval:otfpsi_secret_shared_soft} presents the performance of the secret-shared variants \otfpsiss and \otfpsissb. 

    We observe similar trends as in the instantiation with SilentOT seen in \S\ref{sec:evaluation}. The major differences are slightly faster runtimes at the cost of larger communication. For \otfpsiss, we observe significantly increased communication due to the large number of OTs performed by the protocol.

    \begin{table*}
        \centering
        \caption{Run time and communication of \otfpsi using SoftSpokenOT for querier set size \sizeA, responder set size \sizeB, and dimension $\dimension$ (threshold $\tau = \floor{\dimension/16}$). %
        }
        \label{tab:eval:otfpsi_performance_softspoken}
        \begin{tabular}{rrrrrrrrrrr}
            \toprule
            & & \multicolumn{3}{c}{$\dimension = 127$} & \multicolumn{3}{c}{$\dimension = 511$} & \multicolumn{3}{c}{$\dimension = 8191$}  \\
            \cmidrule(lr){3-5} \cmidrule(lr){6-8} \cmidrule(lr){9-11}
            \sizeA & \sizeB & Gigabit & Slow & Comm & Gigabit & Slow & Comm & Gigabit & Slow & Comm\\
            \midrule
            \num{64} & \num{64} & \qty{0.032}{\second} & \qty{0.744}{\second} & \qty{0.802}{\mebi\byte} & \qty{0.077}{\second} & \qty{0.927}{\second} & \qty{3.079}{\mebi\byte} & \qty{1.242}{\second} & \qty{4.312}{\second} & \qty{72.415}{\mebi\byte} \\
            \num{256} & \num{256} & \qty{0.258}{\second} & \qty{1.432}{\second} & \qty{11.823}{\mebi\byte} & \qty{0.969}{\second} & \qty{2.963}{\second} & \qty{46.017}{\mebi\byte} & \qty{24.523}{\second} & \qty{48.793}{\second} & \qty{1.084}{\gibi\byte} \\
            \num[round-mode=none,exponent-mode=input]{1024} & \num[round-mode=none,exponent-mode=input]{1024} & \qty{3.844}{\second} & \qty{8.724}{\second} & \qty{185.911}{\mebi\byte} & \qty{15.111}{\second} & \qty{30.594}{\second} & \qty{724.021}{\mebi\byte} & \qty{417.561}{\second} & \qty{899.737}{\second} & \qty{17.162}{\gibi\byte} \\
            \num[round-mode=none,exponent-mode=input]{4096} & \num[round-mode=none,exponent-mode=input]{4096} & \qty{61.012}{\second} & \qty{122.616}{\second} & \qty{2.893}{\gibi\byte} & \qty{240.478}{\second} & \qty{477.358}{\second} & \qty{11.266}{\gibi\byte} & \qty[round-precision=4]{6744.221}{\second} & \qty[round-precision=5]{15085.285}{\second} & \qty{273.844}{\gibi\byte} \\
            \midrule
            \num{1} & \num[round-mode=none,exponent-mode=input]{16384} & \qty{0.075}{\second} & \qty{0.912}{\second} & \qty{2.908}{\mebi\byte} & \qty{0.241}{\second} & \qty{1.303}{\second} & \qty{11.276}{\mebi\byte} & \qty{4.583}{\second} & \qty{13.680}{\second} & \qty{273.735}{\mebi\byte} \\
            \num{1} & \num[round-mode=none,exponent-mode=input]{131072} & \qty{0.466}{\second} & \qty{1.841}{\second} & \qty{23.130}{\mebi\byte} & \qty{1.791}{\second} & \qty{4.860}{\second} & \qty{90.026}{\mebi\byte} & \qty{36.457}{\second} & \qty{103.159}{\second} & \qty{2.138}{\gibi\byte} \\
            \num{1} & \num[round-mode=none,exponent-mode=input]{524288} & \qty{1.833}{\second} & \qty{4.702}{\second} & \qty{92.466}{\mebi\byte} & \qty{7.129}{\second} & \qty{15.543}{\second} & \qty{360.029}{\mebi\byte} & \qty{145.812}{\second} & \qty{367.415}{\second} & \qty{8.550}{\gibi\byte} \\
            \num{1} & \num[round-mode=none,exponent-mode=input]{1048576} & \qty{3.635}{\second} & \qty{8.379}{\second} & \qty{184.912}{\mebi\byte} & \qty{14.270}{\second} & \qty{30.429}{\second} & \qty{720.030}{\mebi\byte} & \qty{290.953}{\second} & \qty{739.095}{\second} & \qty{17.100}{\gibi\byte} \\
            \bottomrule
        \end{tabular}
    \end{table*}

    \begin{table*}
        \centering
        \caption{Run time and communication of plaintext \otfpsi and secret-shared \otfpsiss and \otfpsissb with SoftSpokenOT for querier set size \sizeA and responder set size \sizeB (dimension $\dimension = 511$, threshold $\tau = \floor{\dimension/16} = 31$). }

        \label{tab:eval:otfpsi_secret_shared_soft}
        \begin{tabular}{rrrrrrrrrrr}
            \toprule
            &  & \multicolumn{3}{c}{\otfpsi} & \multicolumn{3}{c}{\otfpsiss} & \multicolumn{3}{c}{\otfpsissb} \\
            \cmidrule(lr){3-5} \cmidrule(lr){6-8} \cmidrule(lr){9-11}
            \sizeA & \sizeB & Gigabit & Slow & Comm & Gigabit & Slow & Comm & Gigabit & Slow & Comm \\
            \midrule
            \num{64} & \num{64} & \qty{0.077}{\second} & \qty{0.927}{\second} & \qty{3.079}{\mebi\byte} & \qty{0.240}{\second} & \qty{1.673}{\second} & \qty{18.798}{\mebi\byte} & \qty{0.194}{\second} & \qty{1.238}{\second} & \qty{7.897}{\mebi\byte} \\
            \num[round-mode=none]{256} & \num[round-mode=none]{256} & \qty{0.969}{\second} & \qty{2.963}{\second} & \qty{46.017}{\mebi\byte} & \qty{3.383}{\second} & \qty{13.635}{\second} & \qty{300.519}{\mebi\byte} & \qty{2.576}{\second} & \qty{5.379}{\second} & \qty{120.017}{\mebi\byte} \\
            \num[round-mode=none,exponent-mode=input]{1024} & \num[round-mode=none,exponent-mode=input]{1024} & \qty{15.111}{\second} & \qty{30.594}{\second} & \qty{724.021}{\mebi\byte} & \qty{51.001}{\second} & \qty{197.964}{\second} & \qty{4.695}{\gibi\byte} & \qty{40.720}{\second} & \qty{87.882}{\second} & \qty{1.852}{\gibi\byte} \\
            \num[round-mode=none,exponent-mode=input]{4096} & \num[round-mode=none,exponent-mode=input]{4096} & \qty{240.478}{\second} & \qty{477.358}{\second} & \qty{11.266}{\gibi\byte} & \qty{806.894}{\second} & \qty{3140.775}{\second} & \qty{75.125}{\gibi\byte} & \qty{649.118}{\second} & \qty{1466.105}{\second} & \qty{29.531}{\gibi\byte} \\
            \midrule
            \num{1} & \num[round-mode=none,exponent-mode=input]{16384} & \qty{0.241}{\second} & \qty{1.303}{\second} & \qty{11.276}{\mebi\byte} & \qty{0.897}{\second} & \qty{4.112}{\second} & \qty{75.142}{\mebi\byte} & \multicolumn{3}{c}{\multirow{3}{*}{\makecell{\emph{Batching not applicable}\\ \emph{for $\sizeA=1$}}}}  \\
            \num{1} & \num[round-mode=none,exponent-mode=input]{131072} & \qty{1.791}{\second} & \qty{4.860}{\second} & \qty{90.026}{\mebi\byte} & \qty{6.734}{\second} & \qty{25.997}{\second} & \qty{601.018}{\mebi\byte} &  &  &  \\
            \num{1} & \num[round-mode=none,exponent-mode=input]{524288} & \qty{7.129}{\second} & \qty{15.543}{\second} & \qty{360.029}{\mebi\byte} & \qty{26.775}{\second} & \qty{100.171}{\second} & \qty{2.348}{\gibi\byte} &  &  &  \\
            \bottomrule
        \end{tabular}
    \end{table*}

    \subsection{Comparison with Prior FPSI Protocols}\label{ap:compRW_FPSI}

    \subsubsection{FLPSI}\label{ap:comp_FLPSI} 

    The FLPSI protocol \cite{Uzun2021} approximates the Hamming distance using a sub-sampling approach: For a bit string \elementA, it generates $T$ sub-samples where each sub-sample consists of $L$ bits of \elementA at positions determined by fixed random masks. Two bit strings are considered to match if they have at least $t$ sub-samples in common. In practice, the complexity of their protocol restricts the number of sub-samples $T$ as well as the threshold $t$.

    We evaluate the accuracy of this approach and find it is not sufficient for our application. For this, we use the same parameters as the authors ($\dimension = 256$, $T = 64$, $t=2$, $\threshold = 25$) to sample random pairs of bit strings and compare them using the sub-sampling approach. We observe that \qty{0,0015}{\percent} of pairs are falsely classified as matching even though their Hamming distance is greater than \threshold.

    When comparing one bit string to a database of \num[exponent-mode=input]{100000}, we therefore expect that $1.5$ of the individual comparisons are falsely positive. This effect leads to a substantial overall false-positive rate since almost every queried record will be considered a duplicate of one (or multiple) records in the database, violating \reqlinky{fpr}. Due to this sub-sampling technique, FLPSI is unable to provide the necessary accuracy for humanitarian deduplication.

    \subsubsection{Fmap-FPSI}\label{ap:comp_FmapFPSI}
    We now discuss Fmap-FPSI \cite{Gao2024} in more detail.
    The Fmap primitive relies on the key assumption that each element in the receiver's set has $\threshold +1$ unique components: i.e., for each element, there are $\threshold+1$ dimensions where all other elements have a different value.
    This assumption is very restrictive as it: (i) constricts the threshold in relation to the dimension (e.g., their implementation requires $\dimension > 8(\threshold+1)$), and (ii) limits the maximum receiver set size.
    For example, for bit strings of length $\dimension = 512$ and threshold $\threshold = 128$ and without pre-processing, the largest set fulfilling this assumption has cardinality 3. The authors suggest packing multiple bits into one dimension of a lower-dimensional integer vector.
    For our parameters, we can pack at most three bits into one component (otherwise, the dimension would be less than $\threshold+1$). 

    When packing two or three bits, the assumption cannot hold for any sets of meaningful size. To illustrate this, we sample 16 random bit strings of size $\dimension = 512$ and measure the probability that another random bit string has $\threshold +1 = 129$ unique dimensions with regard to all 16 vectors in the set. When packing three bits, this probability is \num{1.918e-82} and \num{2.163e-183} when packing 2 bits. This illustrates that, in our setting with a relatively high threshold \threshold, we cannot rely on the assumption Fmap-FPSI is based on.

    \begin{table}
        \centering
        \caption{Run time of Fmap-FPSI and \otfpsi with SilentOT by set size $n =\sizeA = \sizeB$ ($\dimension=128$, $\threshold=4$, \qty{10}{\gibi\bit\per\second}, \qty{.02}{\milli\second} latency)}
        \label{tab:eval:comp_fmap_fpsi_paper_silent}
        \begin{tabular}{rrrrrr}
            \toprule
            & \multicolumn{3}{c}{Fmap-FPSI} & \multicolumn{2}{c}{\otfpsi} \\
            \cmidrule(lr){2-4} \cmidrule(lr){5-6}
            $n$ & Online & Total & Comm & Total & Comm \\
            \midrule
            256 & \qty{2.179997}{\second} & \qty{100.288}{\second} & \qty{91.889}{\mebi\byte} & {\bfseries \qty{0.301}{\second}} & \qty{9.219}{\mebi\byte} \\
            1024 & \qty{8.792}{\second} & \qty{401.892}{\second} & \qty{367.529}{\mebi\byte} & {\bfseries \qty{4.542}{\second}} & \qty{145.324}{\mebi\byte} \\
            4096 & {\bfseries \qty{35.390}{\second}} & \qty{1617.623}{\second} & \qty{1,43562793}{\gibi\byte} & \qty{72.073}{\second} & \qty{2.268}{\mebi\byte} \\
            \bottomrule
        \end{tabular}
    \end{table}

    To show the efficiency of \otfpsi, we still compare its performance to Fmap-FPSI. Fmap-FPSI consists of an offline phase that may be re-used as long as the receiver set does not change, and an online phase. We measure both online and offline phases to allow for a fair comparison.  
    \Cref{tab:eval:comp_fmap_fpsi_paper_silent} compares both protocol at the evaluation parameters used by the authors: For set sizes \num{256} and \num[round-mode=none]{1024}, \otfpsi is faster than the online phase of Fmap-FPSI. For set size \num[round-mode=none]{4096}, this is no longer the case. Yet, the offline phase of Fmap-FPSI is so expensive that we would need 26 protocol iterations with the same receiver set until Fmap-FPSI would be faster than \otfpsi in total.

    As \otfpsi scales quadratically in the set size it can be slower than Fmap-FPSI for large sets. However, Fmap-FPSI achieves this linear scaling by relying on a restrictive input assumption and by requiring an offline phase that is concretely expensive. \otfpsi works on any input and our experiments show that it scales better to higher dimensions and thresholds.

    \begin{table*}
        \centering
        \caption{Comparison of the evaluation environments of existing \ac{FPSI} protocols}
        \begin{threeparttable}
            \begin{tabular}{@{}llp{3.1cm}llp{3.2cm}p{2.75cm}@{}}
                \toprule
                & Year & Machine/CPU Type & Cores/Threads & RAM & Network Setting & Parallelism \\
                \midrule
                FLPSI \cite{Uzun2021} & 2021 & Azure F72s\_v2 (Intel Xeon Platinum 8168) & \num{72} vCPUs & \qty{144}{\giga\byte} & Not relevant for comparison\tnote{a} & Online: single-threaded,\tnote{b} offline: unspecified \\
                DA-PSI \cite{Chakraborti2023} & 2023 & \qty{2}{\x} AWS EC2 t2.xlarge\tnote{c} & 4 vCPUs & \qty{16}{\giga\byte} & Real network with \num{320}-\qty{480}{\mega\bit\per\second} and unspecified latency & Unspecified \\
                Approx-PSI \cite{Chongchitmate2024} & 2024 & Unspecified & \num{8} vCPUs & \qty{8}{\giga\byte} & Unspecified LAN and \qty{480}{\mega\bit\per\second} with unspecified latency & Single-threaded \\
                Fmap-FPSI\tnote{d}{} \cite{Gao2024} & 2024 &Intel Xeon Gold 6330 & Unspecified & \qty{256}{\giga\byte} & \qty{10}{\giga\bit\per\second}, \qty{0.02}{\milli\second} latency & Unspecified\tnote{e}\\
                PE-FPSI \cite{Blass2025} & 2025 & AWS EC2 c7i.metal-48xl (Intel Sapphire Rapids 8488C \cite{AWSInstances}) & \num{192} vCPUs \cite{AWSInstances} & \qty{384}{\giga\byte} \cite{AWSInstances} & Unlimited & Unspecified \\
                \midrule
                \textbf{Ours.} & 2025 &Google Cloud c4d-standard-8 (AMD EPYC Turin) & 4 vCPUs\tnote{f} & \qty{30}{\giga\byte} & Gigabit (\qty{1}{\giga\bit\per\second}, \qty{0.5}{\milli\second} latency) and slow (\qty{250}{\mega\bit\per\second}, \qty{20}{\milli\second} latency) & Single-threaded \\
                \bottomrule
            \end{tabular}
            \begin{tablenotes}
                \item[a] We only reproduce their computation costs (excluding communication).
                \item[b] For comparison, we only use their single-threaded results.
                \item[c] The t2.xlarge instances are burstable, i.e. do not provide constant CPU perfornamce over time \cite{AWSInstances}.
                \item[d] We run their code in our environment and only provide their evaluation environment for reference. 
                \item[e] The published source code is single-threaded.
                \item[f] Configured to one vCPU per core. 
            \end{tablenotes}
        \end{threeparttable}
        \label{tab:eval_rw_environments}
    \end{table*}

    \subsection{Comparison to Funshade}
    \label{ap:funshade}

    Funshade \cite{Ibarrondo2023} privately computes distance metrics between two (integer) vectors and privately compare the distance to a threshold. The protocol uses two compute parties and two data holders (that may be different from the compute parties), which is a similar system model to secret-shared FPSI in \name. In the following, we illustrate the limitations of the Funshade protocol.

    Funshade utilizes $\Pi$-sharing, a variant of arithmetic secret sharing that enables one online multiplication without online communication. This is achieved by pre-generating Beaver triplets offline and building the $\Pi$-shares based on the Beaver triplets. After computing secret-shared distances, Funshade performs a threshold comparison using Function Secret Sharing (FSS).

    \descr{Setup. } The setup phase generates Beaver multiplication triplets and FSS keys. In our system model without additional trust assumptions, we can only perform both steps in SMC. However, the authors evaluate Funshade under the assumption that the setup is performed by a trusted third party (TTP). We expect that implementing a setup without a TTP using SMC would increase costs significantly.

    \descr{Cost. } %
    We cannot directly estimate the cost of Funshade for our setting since the authors use parameters tailored for a different distance metric in their evaluation. Further, we were unable to reproduce their results \cite[Table 2]{Ibarrondo2023} using their code.
    Thus, we linearly extrapolate the presented performance numbers to our parameters (for an online query with $\dimension=511$, $n=9$, $\sizeB = 131072$). We estimate that the setup phase using a trusted third party would take  
    \qty{28,4917248}{\second} over a \qty{250}{\mebi\bit\per\second} connection. This ignores the online phase of the protocol and the cost of a secure setup, which we expect to be significantly more expensive. 

    Our protocol \otfpsiss can perform the same query over the same network in just \qty{13.3}{\second}.

    \descr{Multiple Comparisons. } A core benefit of secret-shared FPSI in \name is that a field team can generate secret-shares of a registration once, and the compute nodes can then repeatedly use these shares without the field team's involvement.
    
    For Funshade, it is unclear whether the $\Pi$-shares can be reused for multiple comparisons. The evaluation results \cite[Table 4]{Ibarrondo2023} suggest that shares are reused, while the authors also state that ``fresh randomness is generated for each 1:1 verification''. Fresh shares for each comparison would put a high load on the field teams (violate \reqlinky{local-efficient}), while it is unclear whether share reuse would retain the security guarantees of the protocol.

    \subsection{Evaluation Environments}

    The FPSI works discussed in \cref{sec:evaluation} do not have a publicly available implementation with the exception of Fmap-FPSI. Thus, in our evaluation, we can only build on the evaluation results presented by the authors in their respective papers.  

    \Cref{tab:eval_rw_environments} gives an overview of the evaluation environments used in prior work.

\newpage %

\section{Meta-Review}

The following meta-review was prepared by the program committee for the 2026
IEEE Symposium on Security and Privacy (S\&P) as part of the review process as
detailed in the call for papers.

\subsection{Summary}
This paper addresses the problem of deduplication across humanitarian organizations, where multiple entities need to detect whether the same individual is registered to receive aid from each of them, in order to avoid unnecessary or duplicate assistance. The authors propose a system called xDup that relies on a two-servers FPSI construction based on OT. The authors implement the core FPSI protocol to demonstrate the feasibility of xDup and evaluate its performance.

\subsection{Scientific Contributions}
\begin{itemize}
\item 3. Creates a New Tool to Enable Future Science.
\item 6. Provides a Valuable Step Forward in an Established Field.
\end{itemize}

\subsection{Reasons for Acceptance}
\begin{enumerate}
\item The paper presents a useful tool that tries to addresses an important and concrete real-world problem in the humanitarian sector. By focusing on a realistic deployment scenario, it introduces meaningful practical constraints that guide the protocol design and help bridge a gap in the recent FPSI literature. 
\item The work demonstrates that a well-designed combination of established building blocks (OT, FPSI secret-sharing) can deliver an impactful and practical service.
\item The authors also implement and evaluate their scheme, and their benchmarks show that the proposed construction significantly outperforms existing approaches.
\end{enumerate}

\subsection{Noteworthy Concerns} %
\begin{enumerate} %
\item The paper's central protocol seems a straightforward generalization of the SHADE protocol for obliviously computing hamming distance from OT. Yet, there are some non-trivial observations and adaptions from that SHADE protocol to a FPSI protocol with secret shared inputs.

\item The threat model assumes honest registrations, which may be difficult to guarantee in practice. In real-world humanitarian settings, field teams may face significant challenges in verifying the accuracy of biographical data provided by registrants. If an individual manages to register multiple times under different identities, the protocol may fail to detect such duplication and thus may not fully enforce fairness in aid distribution.
\end{enumerate}

\end{document}